\documentclass[prx,twocolumn,superscriptaddress,secnumarabic,balancelastpage,amsmath,amssymb]{revtex4-1}

\usepackage{graphicx}
\usepackage{latexsym}
\usepackage{amssymb}
\usepackage{amsmath}
\usepackage{amsfonts}
\usepackage{upgreek}
\usepackage{bm}
\usepackage{tcolorbox}
\usepackage{multirow}
\usepackage{enumitem}
\usepackage{color}
\usepackage{hyperref}
\usepackage{verbatim}
\hypersetup{
colorlinks = true,
linkcolor = [rgb]{0.70,0.13,0.13},
citecolor = [rgb]{0.13,0.55,0.13},
urlcolor = [rgb]{0.25, 0.41, 0.88}}
\newcommand{\bra}[1]{\langle #1|}
\newcommand{\ket}[1]{|#1 \rangle}

\newcommand{\Tr}{\operatorname{Tr}}

\renewcommand{\Re}{\operatorname{Re}}
\renewcommand{\Im}{\operatorname{Im}}

\newcommand{\eqnref}[1]{Eq.\,\eqref{#1}}
\newcommand{\figref}[1]{Fig.\,\ref{#1}}
\newcommand{\tabref}[1]{Tab.\,\ref{#1}}

\usepackage{physics}
\newcommand{\rd}{\partial}

\usepackage{bm,bbm}

\let \oldbm \bm
\renewcommand{\vec}[1]{\oldbm{#1}}

\def\bk{{\vec k}}

\def\bm{{\vec m}}

\def\tr{\mathop{\mathrm{tr}}}

\usepackage{slashed}
\usepackage[normalem]{ulem}

\begin{document}

\title{Topological Correspondence between Hermitian and Non-Hermitian Systems: Anomalous Dynamics}

\author{Jong Yeon Lee}
\affiliation{Department of Physics, Harvard University, Cambridge, Massachusetts 02138, USA}

\author{Junyeong Ahn}
\affiliation{Department of Physics and Astronomy, Seoul National University, Seoul 08826, Korea}
\affiliation{Center for Correlated Electron Systems, Institute for Basic Science (IBS), Seoul 08826, Korea}
\affiliation{Center for Theoretical Physics (CTP), Seoul National University, Seoul 08826, Korea}

\author{Hengyun Zhou}
\affiliation{Department of Physics, Harvard University, Cambridge, Massachusetts 02138, USA}

\author{Ashvin Vishwanath}
\affiliation{Department of Physics, Harvard University, Cambridge, Massachusetts 02138, USA}

\date{\today}
\begin{abstract}
The hallmark of symmetry-protected topological (SPT) phases is the existence of anomalous boundary states, which can only be realized with the corresponding bulk system.
In this work, we show that for every Hermitian anomalous boundary mode of the ten Altland-Zirnbauer classes, a non-Hermitian counterpart can be constructed, whose long time dynamics provides a realization of the  anomalous boundary state. We prove that the non-Hermitian counterpart is characterized by a point-gap topological invariant, and furthermore, that the invariant exactly matches that of the corresponding Hermitian anomalous boundary mode. We  thus establish a correspondence between the topological classifications of $(d+1)$-dimensional gapped Hermitian systems and $d$-dimensional point-gapped non-Hermitian systems. We illustrate this general result with a number of examples in different dimensions. 
This work provides a new perspective on point-gap topological invariants in non-Hermitian systems.

\end{abstract}
\maketitle
\textit{Introduction---} In the last few decades, topology has emerged as a central theme in the study of condensed matter physics. The interplay of symmetry and topology has led to a wide variety of interesting phenomena, most notably that of symmetry-protected topological phases (SPTs)~\cite{Kitaev2009, Schnyder2008, senthil2015symmetry,chiu2016classification}. One of the key physical signatures of SPTs are their anomalous boundary states, which can only be realized as $d$-dimensional boundary states of a $(d+1)$-dimensional topological bulk, and cannot appear in a $d$-dimensional bulk model.  

\begin{figure}
 \hspace{-5pt} \includegraphics[width=0.44\textwidth]{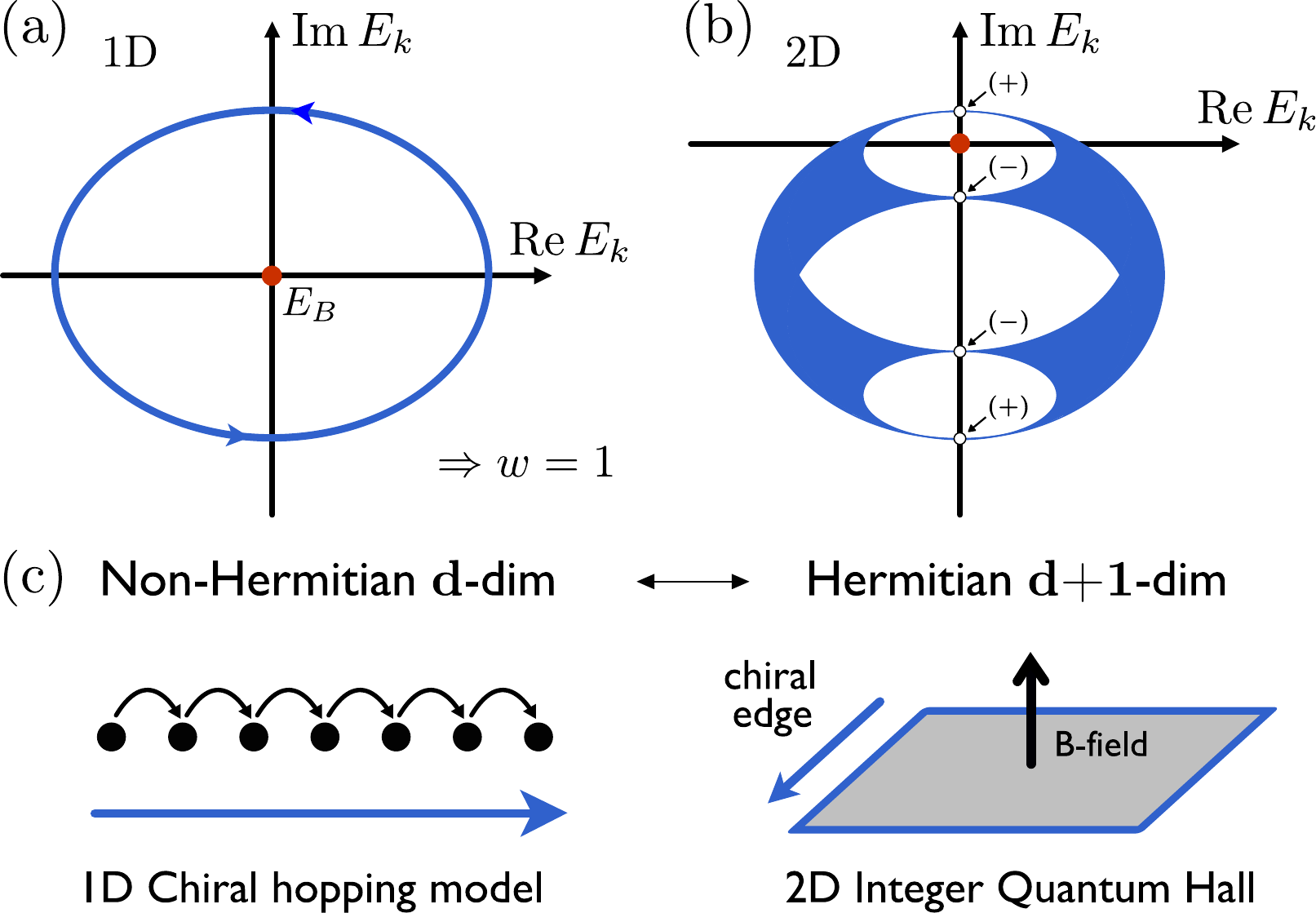}
\caption{ \label{fig:1D_model} (a) Dispersion for the 1D chiral hopping model  in the complex plane. The topological invariant $w$ is defined with respect to $E_B$ as in Eq.~\eqref{eq:1D_invariant}. Depending on the value of $E_B$ and other parameters, $w$ can differ. In this case, $w = 1$. (b) Dispersion for the 2D model in \eqnref{eq:2D_CTI} that resembles the surface of a 3D chiral topological insulator. Here, $\gamma_{\bk} = 2\cos k_x + \cos k_y$, $b_{1,\bk} = \sin k_x$, $b_{2,\bk} = \sin k_y$, $b_{3,\bk} = 0$. Each white dot represents a Dirac cone with $\pm$ chirality. In this case, depending on $E_B$, the topological invariant can be $1,0,-1$.  (c) The 1D system characterized by $w \in \mathbb{Z}$ in the chiral hopping model corresponds to the edge of a 2D system characterized by an integer quantum Hall state with Chern number $n = w \in \mathbb{Z}$.}
\end{figure}

Recently, the study of topological phenomena has also been extended to non-Hermitian systems~\cite{martinez2018topological,hatano1996localization,rudner2009topological,esaki2011edge,yuce2015topological,lee2016anomalous,leykam2017edge,xu2017weyl,shen2018topological,gong2018topological,lieu2018topological,lieu2018topological1,kawabata2018anomalous,kawabata2019topological,kawabata2019non,carlstrom2018exceptional,takata2018photonic,moors2019disorder,zyuzin2018flat,kunst2019non,yang2019non,zhou2019exceptional,budich2019symmetry,okugawa2019topological,dembowski2001experimental,poli2015selective,zeuner2015observation,weimann2016topologically,zhou2018observation,cerjan2019experimental,xiong2018why,kunst2018biorthogonal,yao2018edge,yao2018non,lee2019anatomy,borgnia2019non,zirnstein2019bulk,herviou2019defining,ge2019topological,liu2019second,longhi2019,WuDefect2019,YuceSkin2019,YamamotoNHSF2019,DengSSH2019,Bliokh2019topological,song2019non}, 
which are naturally realized in classical optical systems with gain and loss~\cite{el-ganainy2018non,konotop2016nonlinear,cao2015dielectric,doppler2016dynamically,xu2016topological,zhou2018observation,lapp2019engineering}, superconducting vortices~\cite{hatano1996localization}, ring neural networks~\cite{amir2016non}, bosonic  superconducting systems~\cite{lieu2018topological,mcdonald2018phase}, or magnon band structures~\cite{YML_magnon2018}, and has also been {proposed} to be relevant to electronic systems with finite quasiparticle lifetime~\cite{kozii2017non,shen2018topological,zyuzin2018flat}. In particular, with a suitable generalization of the gap condition~\cite{gong2018topological}, SPTs can be generalized to the non-Hermitian setting~\cite{zhou2019exceptional,okugawa2019topological,budich2019symmetry,zhou2019periodic,kawabata2018symmetry}, and the well-known ten-fold way classification for non-interacting fermionic topological phases under the Altland-Zirnbauer (AZ) symmetry classes~\cite{altland1997nonstandard,Kitaev2009,ryu2010topological} can be extended to the 38 non-Hermitian Bernard-LeClair (BL) symmetry classes~\cite{Bernard2001,sato2012time,lieu2018topological,zhou2019periodic,kawabata2018symmetry}. %
Interestingly, the classification of non-Hermitian SPTs also exhibits a periodic structure similar to Hermitian systems~\cite{gong2018topological,zhou2019periodic,kawabata2018symmetry}, both in symmetry class and spatial dimension, and certain characteristics of the 1D non-Hermitian models studied are reminiscent of boundary states of 2D Hermitian models with related symmetries~\cite{DeMarco2018,zhou2019periodic}.
As an example, the boundary of a 2D quantum Hall system hosts anomalous chiral edge states, which bears some resemblance to the 1D non-Hermitian chiral hopping model, as shown in Fig.~\ref{fig:1D_model}(c).
This raises the question of whether there exists a more general correspondence between the anomalous boundary states of a Hermitian system, and the dynamics of a corresponding non-Hermitian system with one dimension lower.

In this Letter, we establish a correspondence between the {ten-fold-way topological classification} of non-interacting Hermitian systems in $(d+1)$ dimensions and {the point-gap topology of certain} non-Hermitian systems in $d$ dimensions, and describe how this gives a possible interpretation to the long-time dynamics of non-Hermitian models as a \emph{dynamical anomaly}, in direct relation to the anomalous boundary physics of Hermitian systems.  We motivate this by introducing a 1D chiral hopping non-Hermitian model and examining the relation between non-Hermitian band topology and anomalous chiral modes in the long-time limit. We then generalize this to other symmetry classes, and prove the above correspondence in both the topological classification as well as the explicit realization of anomalous dynamics.

\textit{Emergence of chiral fermions in a 1D non-Hermitian system---}
We start by considering an example to motivate and illustrate the main idea of the correspondence. Consider the following single-band non-Hermitian Hamiltonian in one dimension~\cite{hatano1996localization,gong2018topological}:
\begin{equation}\label{eq:1D_model}
H = \sum_{r} \qty( t_L c^\dagger_r c_{r+1} + t_R c^\dagger_{r+1} c_{r} ),
\end{equation}
where $t_R \neq t_L$. Under the Fourier transformation $c_r = \sum_k c_k e^{ik r}$, the $k$-space Hamiltonian is given by $H_k = (t_L + t_R) \cos k + i (t_L - t_R) \sin k$. Thus, the energy dispersion $E_k$ forms an ellipse in the complex energy plane [\figref{fig:1D_model}(a)]. For a positive (negative) $t_L - t_R$, the band winds around the origin in the counterclockwise (clockwise) direction. The group velocity $v_k$ of a wave packet centered at {$k$} is given by~\cite{gong2018topological}
\begin{eqnarray}
v_k &=& \Re \frac{1}{\hbar} \frac{\rd E_k}{\rd k} = - (t_L + t_R) \sin k,
\end{eqnarray}
since the imaginary part of $\rd_k E_k$ does not affect the propagation velocity of the wave packet.

On top of this, there is an additional ingredient that influences the dynamics of a non-Hermitian system, the imaginary part of the energy which causes certain eigenstates to grow or decay with $\Im E_k = (t_L - t_R) \sin k$.
If we inspect the dynamics at real energy near zero, then there may exist two modes: left and right propagating modes with $k=\pm \pi/2$. While the left-propagating mode has a positive $\Im E_k$, the right-propagating mode has a negative $\Im E_k$. Therefore, if we excite the system with a frequency $\omega \sim 0$, generically both counter-propagating modes will be excited, but the right-propagating mode will die out after a timescale $\tau_0 \gg \hbar / \Im E_k$. At long times, we will thus observe chiral dynamics in the system with only a left-propagating mode, a scenario which cannot be realized in any 1D Hermitian system.

\begin{figure}
 \hspace{-5pt} \includegraphics[width=0.49\textwidth]{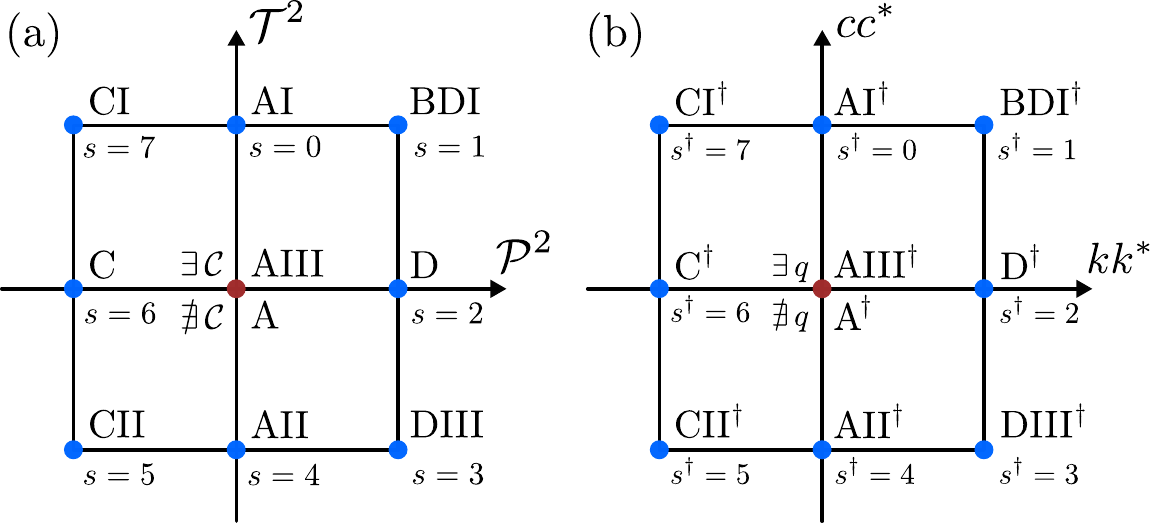}
\caption{\label{fig:AZclasses}
Diagrams defining (a) Hermitian classes $s$ (b) Non Hermitian real classes $s^\dagger$ (AZ$^\dagger$). Hermitian AZ classes are defined by time reversal $\cal T$, particle-hole $\cal P$ and chiral $\cal C$ symmetries. Similarly, non-Hermitian AZ$^\dagger$ classes are defined by $K$, $C$ and $Q$-type symmetries. Real (complex) classes are given by blue (red) dots. There are two complex classes ($s=0,1$) depending on the absence or presence of ${\cal C}$ or $q$.
}
\end{figure}

\begin{table}[t]
\begin{tabular}{|c|c|c|c|c|}
\hline
\,  { ${\cal M}$} \, & \, $\pi_0({\cal M})$ \,	&\,\,AZ class \,\,	& \,\, NH AZ$^{\dagger}$ \,\,    &\,\,NH AZ \,\,	\\
\hline
${\cal C}_0$  & $\mathbb{Z}$		&A 	(3)		&AIII$^\dagger$ (3)	&AIII (3)    \\
${\cal C}_1$    & 0 		&AIII (4)	&A$^\dagger$ (1)			&A (1)				\\
\hline
${\cal R}_0$  & $\mathbb{Z}$ 		&AI (14)	&BDI$^{\dagger}$ (14)    &CI (21)	\\
${\cal R}_1$ 	& $\mathbb{Z}_2$	&BDI (22)	&D$^{\dagger}$ (34)    &AI (34)			\\
${\cal R}_2$    & $\mathbb{Z}_2$		&D (16)	    &DIII$^{\dagger}$	(19)    &BDI (16)	\\
${\cal R}_3$    &   0		&DIII (27)	&AII$^{\dagger}$ (7)    &D (8)			\\
${\cal R}_4$    & $\mathbb{Z}$		&AII (15)	&CII$^{\dagger}$	(15)    &DIII (20)	\\
${\cal R}_5$    & 0 		&CII (23)	&C$^{\dagger}$ (35)	&AII (35)	\\
${\cal R}_6$    & 0		&C (17)	    &CI$^{\dagger}$ (18)    &CII (17)		\\
${\cal R}_7$    & 0		&CI (26)	&AI$^{\dagger}$ (6)    &C (9) \\	
\hline
\end{tabular}
\caption{\label{tab:correspondence} Classifying spaces ${\cal M}$ for Hermitian and non-Hermitian Altland-Zirnbauer (AZ) classes in $0$-dimensional systems. The symmetry classes are defined in Fig.~\ref{fig:AZclasses}. For a general $d$-dimensional system, the classifying space shifts as ${\cal R}_s \mapsto {\cal R}_{s-d}$. Thus, the classification of $d$-dimensional Hermitian class $s$ is equivalent to that of $(d+1)$-dimensional NH $s^\dagger$ and NH ($s-2$) classes. The numbers in the parenthesis show the label used in Ref.~\cite{zhou2019periodic, SM}.}
\end{table}

\textit{Non-Hermitian Topology with Complex Point Gap---} The above chiral dynamics can be directly connected to the topological properties of the non-Hermitian band structure. Here, band topology is defined by the complex point energy gap constraint: for a given complex base energy $E_B$, two band structures are topologically equivalent if and only if one can be deformed to the other without crossing $E_B$ during the deformation~\cite{gong2018topological, zhou2019periodic, kawabata2018symmetry}.
In this context, $\Re E_B$ and $\Im E_B$ are the real energy window and overall loss/gain level we are interested in, respectively. {Note that the choice of point-gap non-Hermitian topology here}, instead of line gaps~\cite{kawabata2018symmetry} or band separation~\cite{shen2018topological}, plays an important role in establishing the correspondence.

For the above model, the explicit topological invariant $w \in \mathbb{Z}$ is given by 
\begin{equation} \label{eq:1D_invariant}
    w = \int_{-\pi}^{\pi} \frac{dk}{2\pi i} \rd_k \ln (E_k-E_B),
\end{equation}
which is nothing but the winding number of $E_k$ around the base point $E_B$ [\figref{fig:1D_model}(a,b)].
As a consequence, a nontrivial winding number $w$ implies the existence of modes at $\Re E_B$, some with imaginary part above $\Im E_B$ and others below $\Im E_B$. For the base point choice in Fig.~\ref{fig:1D_model}(a), $w=1$ for $t_L > t_R$ and $w=-1$ for $t_L < t_R$. Examining the expressions for the group velocity and imaginary part in the preceding section, we see that $w$ directly corresponds to the number of left-propagating modes minus the number of right-propagating modes, with imaginary part above $\Im E_B$, which in turn characterizes the total chirality of long time dynamics. 
Therefore, the non-Hermitian topological invariant $w$ indeed captures the anomalous dynamics of this model.

{\it Hermitian-Non-Hermitian correspondence---}
The preceding 1D chiral hopping model hints at a nontrivial connection between non-Hermitian band topology and anomalous dynamics. In particular, the model is reminiscent of the anomalous edge states in a two-dimensional integer quantum Hall state, where the topological invariant $n \in \mathbb{Z}$ characterizes the number of chiral edge modes [\figref{fig:1D_model}(c)]. A similar correspondence has also been pointed out by some of the authors~\cite{zhou2019periodic} in higher dimensions. Below, we will make this correspondence more rigorous, proving the following general statement:

\vspace{0.1in}
\begin{tcolorbox}
\noindent {\bf Proposition} {For a given {$d$-dimensional} anomalous boundary state of a $(d+1)$-dimensional Hermitian system in a symmetry class $s$, characterized by a topological invariant $n$, there exists a corresponding $d$-dimensional non-Hermitian topological system on a closed manifold in the class $s^{\dagger}$ (and $s-2$) that realizes the same anomalous physics as its long time dynamics, characterized by a non-Hermitian topological invariant $n$ defined with respect to a certain $E_B$.}
\end{tcolorbox}
\vspace{0.1in}

\noindent {Note that since the anomalous boundary theory of the Hermitian system in one higher dimension is defined on a closed manifold, the corresponding non-Hermitian system is also defined on a closed manifold, thus avoiding the non-Hermitian skin effect \cite{yao2018edge,yao2018non,martinez2018topological,xiong2018why,lee2019anatomy,kunst2018biorthogonal,borgnia2019non}.}
The exact correspondence is summarized in Tab.~\ref{tab:correspondence}. To understand the table, we need to introduce the following Bernard-LeClair non-Hermitian symmetries~\cite{Bernard2001}, which generalize the AZ symmetry classes:
\begin{align}
{H({\bf k})}&=\epsilon_q q{H^\dagger({\bf k})} q^{-1},& q^2&=\mathbb{I}&(Q \textrm{ sym.})\label{eq:symQ}\\
{H(-{\bf k})}&=\epsilon_c c{H^T({\bf k})}c^{-1},& cc^*&=\eta_c \mathbb{I}&(C \textrm{ sym.})\label{eq:symC}\\
{H(-{\bf k})}&=\epsilon_k k{H^*({\bf k})}k^{-1},& kk^*&=\eta_k \mathbb{I}&(K \textrm{ sym.}) \label{eq:symK}
\\
{H({\bf k})}&= -p{H({\bf k})} p^{-1},& p^2&=\mathbb{I}&(P \textrm{ sym.})\label{eq:symP}
\end{align}
where $\epsilon_{\cal O}, \eta_{\cal O} \in \{1,-1\}$.
These give arise to 38 symmetry classes~\cite{zhou2019periodic, kawabata2018symmetry}, containing the famous ten-fold AZ classes (two complex classes $s=0,1$ and eight real classes $s=0,1,...,7$, see Fig.~\ref{fig:AZclasses}(a)) as a special case.
Instead of dealing with all 38 symmetry classes, we will focus on a subset of them, namely the non-Hermitian (NH) AZ$^\dagger$ classes, defined by~\cite{kawabata2018symmetry}:
\begin{align} \label{eq:AZ_dagger}
k H^*({\bf k})k ^{-1}
&=-H(-{\bf k})
&\text{particle-hole } {\cal P}\mapsto K,\\
c H^T({\bf k})c^{-1}
&=H(-{\bf k})
&\text{time-reversal } {\cal T}\mapsto C,
\end{align}
where the Hermitian chiral symmetry ${\cal C}$, given by the composition of ${\cal T}$ and ${\cal P}$, is replaced by a $Q$-type symmetry with $\epsilon_q = -1$, given by the composition of $C$ and $K$.
With these basic symmetries, the complex or real NH classes $s^\dagger$ are defined as in Fig.~\ref{fig:AZclasses}, which show the same ``Bott clock'' structure as the Hermitian case~\cite{ryu2010topological}.

To see why this is the natural extension of Hermitian AZ classes, we examine how these non-Hermitian symmetries affect the eigenvalue spectrum. As discussed in Ref.~\cite{zhou2019periodic,SM}, one can prove that the chosen $C$ and $K$ type symmetries affect the structure of eigenvalues as follows:
\begin{itemize}[topsep=5pt,parsep=0pt,partopsep=0pt,leftmargin=10pt,labelwidth=6pt,labelsep=4pt]
    \item Hermitian systems: ${\cal P}$ guarantees that eigenvalues appear in a positive and negative pair. ${\cal T}$ with ${\cal T}^2 = -1$ guarantees the Kramers degeneracy.
    \item Non-Hermitian systems: $K$ symmetry guarantees that eigenvalues appear in a pair $(\lambda, \epsilon_k \lambda^*)$. In the case of $\epsilon_k = -1$, this corresponds to a pair of complex energies with opposite real part. $C$ symmetry with $c c^* = -1$ guarantees the biorthonormal Kramers degeneracy. 
\end{itemize}
Thus, the spectral consequences of the choice of symmetry in the AZ$^\dagger$ classes are consistent with the Hermitian case, justifying the above generalizations to non-Hermitian systems. Interestingly, {this symmetry correspondence also naturally} arises in the context of non-Hermitian transfer matrices describing the decaying boundary modes of one-dimensional SPTs~\cite{SM}.  

We note that one can also define NH AZ classes by switching the roles of complex conjugation and transpose symmetries:
\begin{align}
c H^T({\bf k}) c^{-1}
&=-H(-{\bf k})
&\text{particle-hole }\mathcal{P}\mapsto C,\\
k H^*({\bf k}) k^{-1}
&=H(-{\bf k})
&\text{ time-reversal }\mathcal{T}\mapsto K.
\end{align}
Furthermore, a mapping between the classifications of the NH classes $s^\dagger$ and $s-2$ can be explicitly constructed, as summarized in Tab.~\ref{tab:correspondence}. For the proof, see the Supplemental Material \cite{SM}.

\textit{Proof Part I. Dimensional Ascension---} {We now move on to prove our main proposition.} First, we prove the equivalence of the classifications of Hermitian AZ and NH AZ$^\dagger$ classes, making use of the fact that the non-Hermitian topology of $H$ with respect to the base point $E_B$ is equivalent to the Hermitian topology of the following doubled Hamiltonian $\bar{H}$~\cite{gong2018topological} with respect to the zero Fermi energy:
\begin{align}
\bar{H}
=
\begin{pmatrix}
0&H - E_B\\
H^{\dagger} - E_B^*&0
\end{pmatrix}.
\end{align}
Without loss of generality, we set $E_B =0$ from now on. $\bar{H}$ should satisfy the corresponding doubled symmetries and an additional chiral symmetry:
\begin{align}
\bar{c}\bar{H}^*({\bf k})\bar{c}^{-1}
&=\bar{H}(-{\bf k}),\notag\\
\bar{k}\bar{H}^*({\bf k})\bar{k}^{-1}
&=-\bar{H}(-{\bf k}),\notag\\
\Sigma\bar{H}({\bf k})\Sigma^{-1}
&=-\bar{H}({\bf k}),
\end{align}
where $\bar{k}=I\otimes k$, $\bar{c}=\sigma_x\otimes c$, $\Sigma=\sigma_z\otimes I$, $\bar{k}\bar{k}^*=\eta_k I$, $\bar{c}\bar{c}^*=\eta_c I$.

Let us start with the doubled Hamiltonian $\bar{H}$ of a $d$-dimensional NH Hamiltonian $H$ in the class $s^{\dagger}$. Following Teo and Kane~\cite{teo2010topological}, we can construct a $(d+1)$-dimensional Hamiltonian by introducing a new momentum-like parameter $-\pi/2\le \theta\le \pi/2$~\footnote{More precisely, $\theta$ is a latitude for this higher dimensional Brillouin zone, which is given by the suspension of the original one},
\begin{align}
H_{d+1}=\cos \theta \bar{H}+\sin \theta \Sigma.
\end{align}
One immediately sees that this $(d+1)$-dimensional Hermitian Hamiltonian belongs to class $s$, since
\begin{align}
\bar{c}H_{d+1}^*({\bf k},\theta)\bar{c}^{-1}
&=H_{d+1}(-{\bf k},-\theta),\notag\\
\bar{k}H_{d+1}^*({\bf k},\theta)\bar{k}^{-1}
&=-H_{d+1}(-{\bf k},-\theta),
\end{align}
with $c$ corresponding to time-reversal and $k$ corresponding to particle-hole in the AZ$^{\dagger}$ class. Note that $\Sigma$ is not a symmetry operator anymore. 
Since $\bar{H}$ and $\Sigma$ anticommute, the gap for $H_{d+1}$ closes if and only if the gap for $\bar{H}$ closes and $\sin\theta=0$. Therefore, the classification problems of the Hermitian class $s$ in $(d+1)$ dimensions and the NH class $s^{\dagger}$ in $d$ dimensions are equivalent. From the mapping between NH AZ and AZ$^\dagger$ classes, further equivalence with the NH class $s-2$ follows.

\textit{Proof Part II. Dynamical Anomaly---} Now that we have established an exact correspondence between Hermitian and non-Hermitian classifications [Tab.~\ref{tab:correspondence}], we turn to investigate the anomalous behavior, and show how a non-Hermitian topological system realizes in its long-time dynamics the anomalous boundary physics of a corresponding Hermitian system. Since anomalous boundary states of Hermitian systems appear as Dirac or Weyl fermions \footnote{\label{footnote:deform}Lorentz symmetry is broken in crystals, so Dirac and Weyl fermions that require four-fold or more degeneracy are not stabilized in higher dimensions. Instead, anomalous boundary states (e.g. surface) appear as symmetry-preserving deformations of Dirac or Weyl fermions, with the same anomaly as the Dirac or Weyl fermions.},  
let us consider a boundary state of a topological band structure characterized by a (positive) unit topological invariant, which is given by the following Dirac (or Weyl) Hamiltonian
\begin{align}
H_{\rm Dirac}({\bf k})
&=k_1 \Gamma_1+\cdots+ k_d\Gamma_d,
\end{align}
where $\Gamma_{i=1,...,d}$ are Hermitian matrices that satisfy the Clifford algebra $\{\Gamma_i,\Gamma_j\}= 2 \delta_{ij} $.
Suppose that $H_{\rm Dirac}$ is in the Hermitian AZ class $s$. Correspondingly, we can construct a NH Hamiltonian in the class $s^{\dagger}$:
\begin{align}
\label{eq:Dirac}
H({\bf k})&=i\gamma({\bf k})+h({\bf k}),  \\
h({\bf k})
&=\sin k_1 \Gamma_1+\cdots+\sin k_d \Gamma_d,\notag\\
\gamma({\bf k})
&=\cos k_1 + \cdots + \cos k_d - m,\notag
\end{align}
with $d-2 < m < d$, so that $\gamma({\bf k})$ is positive at ${\bf k}=0$ and negative at all other time-reversal invariant momenta (TRIM).
Here, type $K$ and $C$ symmetries would imply  $[K,\Gamma_i] = 0$ and $\{C,\Gamma_i\} = 0$ \footnote{$k \Gamma_i^* + \Gamma_i k = 0$ and $c \Gamma_i^T - \Gamma_i c = 0$}.
This Hamiltonian has a finite complex energy gap over the whole Brillouin zone as long as $\gamma({\bf k}) \neq 0$ at TRIMs. Since $H$ is in class $s^{\dagger}$, and type $K$ and $C$ symmetries act in the same way as the usual Hermitian symmetries on the Hermitian component $h({\bf k})$ of $H$, $h({\bf k})$ is in Hermitian class $s$. 
Near a TRIM, $h({\bf k})$ describes a Dirac point. Among the $2^d$ Dirac points at TRIMs, only the Dirac cone at ${\bf k}=0$ survives at long times because only $\gamma(0)$ is positive and all other $\gamma({\rm TRIM})$s are negative. Thus, at long times, the non-Hermitian system we have constructed resembles the single Dirac cone anomalous physics of the Hermitian boundary state.

How can this anomalous physics be associated with the nontrivial topology of a non-Hermitian Hamiltonian? To illustrate this, it is sufficient to show that the topology of the corresponding doubled Hamiltonian $\bar{H}$ is nontrivial:
\begin{align}
\label{eq:doubled_TI}
\bar{H}({\bf k})
&=\tau_x \otimes  h({\bf k})-\tau_y\otimes \gamma({\bf k})\notag\\
&=\sin k_1 \tau_x\otimes\Gamma_1+\cdots+\sin k_d \tau_x\otimes\Gamma_d\notag\\
&\quad-\tau_y(\cos k_1+\cdots+\cos k_d - m),
\end{align}
where $\tau_{x,y,z}$ are Pauli matrices. When $d-2<m<d$, this is the Hamiltonian of the $d$-dimensional topological insulator in class $s$ with an additional chiral symmetry ($\Sigma=\tau_z$). %
To see that this Hamiltonian has a unit topological invariant, we consider deformations from the phase with $m>d$, where $\gamma({\bf k})$ is completely negative over the whole Brillouin zone and the system thus lies in the trivial insulator limit \footnote{Let $f_0=\gamma({\bf k})$ and $f_i=\sin k_i$. Then, $\bar{H}: \textrm{BZ}^d \rightarrow (f_0,...,f_d)$
is homotopic to a trivial constant map $g = (\gamma_0, 0, ...,0)$ with $\gamma_0 < 0$. This can be proven by constructing the following homotopy map
\begin{equation}
    F(\mathbf{k},t) = (1-t) f(\mathbf{k}) + t g(\mathbf{k}),
\end{equation}
which gives a fully gapped Hamiltonian for $(\mathbf{k},t) \in \textrm{BZ}^d \times [0,1]$, where $F(k,0) = f(k)$ and $F(k,1) = g(k)$.}.
To reach the range $d-2<m<d$, the band gap goes through the gap closing at $m=d$ at which the system becomes a semimetal with a single Dirac cone at ${\bf k}=0$. Note that this Dirac cone consists of two copies of the symmetry-protected Dirac cone, which can only be gapped as a pair.
When the sign of the mass term is reversed at $m=d$, the topological invariant changes by a single unit, so the phase $d-2<m<d$ has a unit topological invariant. Therefore, a non-Hermitian system carrying nontrivial band topology is topologically equivalent to Eq.~\eqref{eq:Dirac} that exhibits anomalous dynamics. The correspondence can be easily generalized into an anomalous boundary state with $n > 1$, for example, by using multiple copies of Eq.~\eqref{eq:Dirac}.
Moreover, one can also prove conversely that a non-Hermitian system displaying anomalous dynamics of a corresponding Hermitian system must carry nontrivial band topology (see the Supplemental
Material~\cite{SM}). Therefore, there is indeed a rigorous connection between non-Hermitian band topology and its anomalous dynamics. A similar correspondence between the Hermitian class $s$ and the NH class $s-2$ is shown in the Supplemental
Material~\cite{SM}. 

\textit{Example in 2D and {General Anomalies}---} 
Consider a chiral topological insulator (TI) in 3D {belonging} to the Hermitian AIII class, characterized by a topological invariant $n \in \mathbb{Z}$ representing the net chirality of boundary Dirac cones. The corresponding non-Hermitian system is the NH class AIII$^\dagger$ in 2D with pseudo-Hermiticity given by $q h^\dagger q^{-1} = -h$. The following NH Hamiltonian belongs to NH class AIII$^\dagger$ with $q = \sigma_3$:
\begin{equation}\label{eq:2D_CTI}
    h(\bk) = i \gamma_\bk + b_{1,\bk} \sigma_1 + b_{2,\bk} \sigma_2 + i b_{3,\bk} \sigma_3,
\end{equation}
where $\gamma_\bk, b_{i,\bk}$ are real functions of $\bk = (k_x,k_y)$. In \figref{fig:1D_model}(b), the 2D complex dispersion is drawn for a specific choice of parameters. Here, white dots represent Dirac cones, and out of the four Dirac cones, only the one with positive chirality survives at long times in this case, showing that the model corresponds to the boundary of the $n=1$ 3D chiral TI. 
However, the Dirac cone is not the most general anomalous feature in the non-Hermitian setting for dimension higher than one. 
Under non-Hermitian perturbations, it is well known that Dirac cones deform into an exotic exceptional surface structure~\cite{zhou2019exceptional,budich2019symmetry,okugawa2019topological} in $d \geq 2$ (cf. boundary of TI in $d\geq 4$). Indeed, for $b_{3,\bk} \neq 0$, Dirac cones are deformed to nodal exceptional lines. 
When both $\gamma_\bk$ and $b_{i,\bk}$ are odd functions of $\bk$, the model also belongs to the class AII$^\dagger$ with $c= \sigma_2$, which corresponds to the boundary of three-dimensional topological insulators in class AII. In this case, one can show that the number of Dirac cones above $E_B$ is only equivalent modulo two, which agrees with the corresponding Hermitian physics. For detailed discussions with an explicit model, see the Supplemental Material~\cite{SM}.

\textit{Conclusion and Outlook---} In this Letter, we showed that for a given anomalous boundary of a Hermitian ten-fold class, there is a non-Hermitian bulk system exhibiting the same anomalous dynamics and characterized by a corresponding nontrivial point-gap topology.  
Our work is in contrast to recent works exploring possible bulk-boundary correspondences in non-Hermitian systems~\cite{yao2018edge,yao2018non,martinez2018topological,xiong2018why,lee2019anatomy,kunst2018biorthogonal,borgnia2019non}, 
as we focus only on the bulk physics of  non-Hermitian systems under periodic boundary conditions, a natural choice due to the correspondence with Hermitian anomalous boundary theories. 
In the Hermitian ten-fold way classifications, topologically protected boundary modes result from {multiple bands with non-trivial separations}. On the other hand, non-trivial point-gap topology can be well-defined even for a single band, which cannot give rise to a conventional topologically-protected boundary mode. 
Therefore, instead of point-gap topology, other classes of topology, such as line-gap topology~\cite{kawabata2018symmetry}, where the topological constraint implies separation between bands, may be a more natural setting to generalize the bulk-boundary correspondence to non-Hermitian systems. {Indeed, if this holds, our work may also have interesting extensions to a full correspondence between non-Hermitian point-gap topology and boundary modes of line-gap topology.}
%Indeed, the line-gap topology naturally incorporates the Hermitian ten-fold way. If this holds, our work msay further extend to a full correspondence between non-Hermitian point-gap and line-gap topology.

\acknowledgements

J.Y.L. and A.V. are supported by a Simons Investigator Fellowship and by NSF-DMR 1411343. J.A. is supported by IBS-R009-D1.

\bibliography{main}

\begin{thebibliography}{77}%
\makeatletter
\providecommand \@ifxundefined [1]{%
 \@ifx{#1\undefined}
}%
\providecommand \@ifnum [1]{%
 \ifnum #1\expandafter \@firstoftwo
 \else \expandafter \@secondoftwo
 \fi
}%
\providecommand \@ifx [1]{%
 \ifx #1\expandafter \@firstoftwo
 \else \expandafter \@secondoftwo
 \fi
}%
\providecommand \natexlab [1]{#1}%
\providecommand \enquote  [1]{``#1''}%
\providecommand \bibnamefont  [1]{#1}%
\providecommand \bibfnamefont [1]{#1}%
\providecommand \citenamefont [1]{#1}%
\providecommand \href@noop [0]{\@secondoftwo}%
\providecommand \href [0]{\begingroup \@sanitize@url \@href}%
\providecommand \@href[1]{\@@startlink{#1}\@@href}%
\providecommand \@@href[1]{\endgroup#1\@@endlink}%
\providecommand \@sanitize@url [0]{\catcode `\\12\catcode `\$12\catcode
  `\&12\catcode `\#12\catcode `\^12\catcode `\_12\catcode `\%12\relax}%
\providecommand \@@startlink[1]{}%
\providecommand \@@endlink[0]{}%
\providecommand \url  [0]{\begingroup\@sanitize@url \@url }%
\providecommand \@url [1]{\endgroup\@href {#1}{\urlprefix }}%
\providecommand \urlprefix  [0]{URL }%
\providecommand \Eprint [0]{\href }%
\providecommand \doibase [0]{http://dx.doi.org/}%
\providecommand \selectlanguage [0]{\@gobble}%
\providecommand \bibinfo  [0]{\@secondoftwo}%
\providecommand \bibfield  [0]{\@secondoftwo}%
\providecommand \translation [1]{[#1]}%
\providecommand \BibitemOpen [0]{}%
\providecommand \bibitemStop [0]{}%
\providecommand \bibitemNoStop [0]{.\EOS\space}%
\providecommand \EOS [0]{\spacefactor3000\relax}%
\providecommand \BibitemShut  [1]{\csname bibitem#1\endcsname}%
\let\auto@bib@innerbib\@empty
%</preamble>
\bibitem [{\citenamefont {Kitaev}(2009)}]{Kitaev2009}%
  \BibitemOpen
  \bibfield  {author} {\bibinfo {author} {\bibfnamefont {A.}~\bibnamefont
  {Kitaev}},\ }in\ \href {\doibase 10.1063/1.3149495} {\emph {\bibinfo
  {booktitle} {AIP Conference Proceedings}}},\ Vol.\ \bibinfo {volume} {1134}\
  (\bibinfo  {publisher} {AIP},\ \bibinfo {year} {2009})\ pp.\ \bibinfo {pages}
  {22--30},\ \Eprint {http://arxiv.org/abs/0901.2686} {arXiv:0901.2686}
  \BibitemShut {NoStop}%
\bibitem [{\citenamefont {Schnyder}\ \emph {et~al.}(2008)\citenamefont
  {Schnyder}, \citenamefont {Ryu}, \citenamefont {Furusaki},\ and\
  \citenamefont {Ludwig}}]{Schnyder2008}%
  \BibitemOpen
  \bibfield  {author} {\bibinfo {author} {\bibfnamefont {A.~P.}\ \bibnamefont
  {Schnyder}}, \bibinfo {author} {\bibfnamefont {S.}~\bibnamefont {Ryu}},
  \bibinfo {author} {\bibfnamefont {A.}~\bibnamefont {Furusaki}}, \ and\
  \bibinfo {author} {\bibfnamefont {A.~W.~W.}\ \bibnamefont {Ludwig}},\ }\href
  {\doibase 10.1103/PhysRevB.78.195125} {\bibfield  {journal} {\bibinfo
  {journal} {Phys. Rev. B}\ }\textbf {\bibinfo {volume} {78}},\ \bibinfo
  {pages} {195125} (\bibinfo {year} {2008})}\BibitemShut {NoStop}%
\bibitem [{\citenamefont {Senthil}(2015)}]{senthil2015symmetry}%
  \BibitemOpen
  \bibfield  {author} {\bibinfo {author} {\bibfnamefont {T.}~\bibnamefont
  {Senthil}},\ }\href@noop {} {\bibfield  {journal} {\bibinfo  {journal} {Annu.
  Rev. Condens. Matter Phys.}\ }\textbf {\bibinfo {volume} {6}},\ \bibinfo
  {pages} {299} (\bibinfo {year} {2015})}\BibitemShut {NoStop}%
\bibitem [{\citenamefont {Chiu}\ \emph {et~al.}(2016)\citenamefont {Chiu},
  \citenamefont {Teo}, \citenamefont {Schnyder},\ and\ \citenamefont
  {Ryu}}]{chiu2016classification}%
  \BibitemOpen
  \bibfield  {author} {\bibinfo {author} {\bibfnamefont {C.-K.}\ \bibnamefont
  {Chiu}}, \bibinfo {author} {\bibfnamefont {J.~C.}\ \bibnamefont {Teo}},
  \bibinfo {author} {\bibfnamefont {A.~P.}\ \bibnamefont {Schnyder}}, \ and\
  \bibinfo {author} {\bibfnamefont {S.}~\bibnamefont {Ryu}},\ }\href {\doibase
  10.1103/RevModPhys.88.035005} {\bibfield  {journal} {\bibinfo  {journal}
  {Reviews of Modern Physics}\ }\textbf {\bibinfo {volume} {88}},\ \bibinfo
  {pages} {035005} (\bibinfo {year} {2016})}\BibitemShut {NoStop}%
\bibitem [{\citenamefont {{Martinez Alvarez}}\ \emph
  {et~al.}(2018)\citenamefont {{Martinez Alvarez}}, \citenamefont {{Barrios
  Vargas}}, \citenamefont {Berdakin},\ and\ \citenamefont {{Foa
  Torres}}}]{martinez2018topological}%
  \BibitemOpen
  \bibfield  {author} {\bibinfo {author} {\bibfnamefont {V.~M.}\ \bibnamefont
  {{Martinez Alvarez}}}, \bibinfo {author} {\bibfnamefont {J.~E.}\ \bibnamefont
  {{Barrios Vargas}}}, \bibinfo {author} {\bibfnamefont {M.}~\bibnamefont
  {Berdakin}}, \ and\ \bibinfo {author} {\bibfnamefont {L.~E.~F.}\ \bibnamefont
  {{Foa Torres}}},\ }\href {\doibase 10.1140/epjst/e2018-800091-5} {\bibfield
  {journal} {\bibinfo  {journal} {The European Physical Journal Special
  Topics}\ }\textbf {\bibinfo {volume} {227}},\ \bibinfo {pages} {1295}
  (\bibinfo {year} {2018})}\BibitemShut {NoStop}%
\bibitem [{\citenamefont {Hatano}\ and\ \citenamefont
  {Nelson}(1996)}]{hatano1996localization}%
  \BibitemOpen
  \bibfield  {author} {\bibinfo {author} {\bibfnamefont {N.}~\bibnamefont
  {Hatano}}\ and\ \bibinfo {author} {\bibfnamefont {D.~R.}\ \bibnamefont
  {Nelson}},\ }\href {\doibase 10.1103/PhysRevLett.77.570} {\bibfield
  {journal} {\bibinfo  {journal} {Physical Review Letters}\ }\textbf {\bibinfo
  {volume} {77}},\ \bibinfo {pages} {570} (\bibinfo {year} {1996})}\BibitemShut
  {NoStop}%
\bibitem [{\citenamefont {Rudner}\ and\ \citenamefont
  {Levitov}(2009)}]{rudner2009topological}%
  \BibitemOpen
  \bibfield  {author} {\bibinfo {author} {\bibfnamefont {M.~S.}\ \bibnamefont
  {Rudner}}\ and\ \bibinfo {author} {\bibfnamefont {L.~S.}\ \bibnamefont
  {Levitov}},\ }\href {\doibase 10.1103/PhysRevLett.102.065703} {\bibfield
  {journal} {\bibinfo  {journal} {Physical Review Letters}\ }\textbf {\bibinfo
  {volume} {102}},\ \bibinfo {pages} {065703} (\bibinfo {year}
  {2009})}\BibitemShut {NoStop}%
\bibitem [{\citenamefont {Esaki}\ \emph {et~al.}(2011)\citenamefont {Esaki},
  \citenamefont {Sato}, \citenamefont {Hasebe},\ and\ \citenamefont
  {Kohmoto}}]{esaki2011edge}%
  \BibitemOpen
  \bibfield  {author} {\bibinfo {author} {\bibfnamefont {K.}~\bibnamefont
  {Esaki}}, \bibinfo {author} {\bibfnamefont {M.}~\bibnamefont {Sato}},
  \bibinfo {author} {\bibfnamefont {K.}~\bibnamefont {Hasebe}}, \ and\ \bibinfo
  {author} {\bibfnamefont {M.}~\bibnamefont {Kohmoto}},\ }\href {\doibase
  10.1103/PhysRevB.84.205128} {\bibfield  {journal} {\bibinfo  {journal}
  {Physical Review B}\ }\textbf {\bibinfo {volume} {84}},\ \bibinfo {pages}
  {205128} (\bibinfo {year} {2011})}\BibitemShut {NoStop}%
\bibitem [{\citenamefont {Yuce}(2015)}]{yuce2015topological}%
  \BibitemOpen
  \bibfield  {author} {\bibinfo {author} {\bibfnamefont {C.}~\bibnamefont
  {Yuce}},\ }\href {\doibase 10.1016/J.PHYSLETA.2015.02.011} {\bibfield
  {journal} {\bibinfo  {journal} {Physics Letters A}\ }\textbf {\bibinfo
  {volume} {379}},\ \bibinfo {pages} {1213} (\bibinfo {year}
  {2015})}\BibitemShut {NoStop}%
\bibitem [{\citenamefont {Lee}(2016)}]{lee2016anomalous}%
  \BibitemOpen
  \bibfield  {author} {\bibinfo {author} {\bibfnamefont {T.~E.}\ \bibnamefont
  {Lee}},\ }\href {\doibase 10.1103/PhysRevLett.116.133903} {\bibfield
  {journal} {\bibinfo  {journal} {Physical Review Letters}\ }\textbf {\bibinfo
  {volume} {116}},\ \bibinfo {pages} {133903} (\bibinfo {year}
  {2016})}\BibitemShut {NoStop}%
\bibitem [{\citenamefont {Leykam}\ \emph {et~al.}(2017)\citenamefont {Leykam},
  \citenamefont {Bliokh}, \citenamefont {Huang}, \citenamefont {Chong},\ and\
  \citenamefont {Nori}}]{leykam2017edge}%
  \BibitemOpen
  \bibfield  {author} {\bibinfo {author} {\bibfnamefont {D.}~\bibnamefont
  {Leykam}}, \bibinfo {author} {\bibfnamefont {K.~Y.}\ \bibnamefont {Bliokh}},
  \bibinfo {author} {\bibfnamefont {C.}~\bibnamefont {Huang}}, \bibinfo
  {author} {\bibfnamefont {Y.~D.}\ \bibnamefont {Chong}}, \ and\ \bibinfo
  {author} {\bibfnamefont {F.}~\bibnamefont {Nori}},\ }\href {\doibase
  10.1103/PhysRevLett.118.040401} {\bibfield  {journal} {\bibinfo  {journal}
  {Physical Review Letters}\ }\textbf {\bibinfo {volume} {118}},\ \bibinfo
  {pages} {040401} (\bibinfo {year} {2017})}\BibitemShut {NoStop}%
\bibitem [{\citenamefont {Xu}\ \emph {et~al.}(2017)\citenamefont {Xu},
  \citenamefont {Wang},\ and\ \citenamefont {Duan}}]{xu2017weyl}%
  \BibitemOpen
  \bibfield  {author} {\bibinfo {author} {\bibfnamefont {Y.}~\bibnamefont
  {Xu}}, \bibinfo {author} {\bibfnamefont {S.~T.}\ \bibnamefont {Wang}}, \ and\
  \bibinfo {author} {\bibfnamefont {L.~M.}\ \bibnamefont {Duan}},\ }\href
  {\doibase 10.1103/PhysRevLett.118.045701} {\bibfield  {journal} {\bibinfo
  {journal} {Physical Review Letters}\ }\textbf {\bibinfo {volume} {118}},\
  \bibinfo {pages} {045701} (\bibinfo {year} {2017})}\BibitemShut {NoStop}%
\bibitem [{\citenamefont {Shen}\ \emph {et~al.}(2018)\citenamefont {Shen},
  \citenamefont {Zhen},\ and\ \citenamefont {Fu}}]{shen2018topological}%
  \BibitemOpen
  \bibfield  {author} {\bibinfo {author} {\bibfnamefont {H.}~\bibnamefont
  {Shen}}, \bibinfo {author} {\bibfnamefont {B.}~\bibnamefont {Zhen}}, \ and\
  \bibinfo {author} {\bibfnamefont {L.}~\bibnamefont {Fu}},\ }\href {\doibase
  10.1103/PhysRevLett.120.146402} {\bibfield  {journal} {\bibinfo  {journal}
  {Physical Review Letters}\ }\textbf {\bibinfo {volume} {120}},\ \bibinfo
  {pages} {146402} (\bibinfo {year} {2018})}\BibitemShut {NoStop}%
\bibitem [{\citenamefont {Gong}\ \emph {et~al.}(2018)\citenamefont {Gong},
  \citenamefont {Ashida}, \citenamefont {Kawabata}, \citenamefont {Takasan},
  \citenamefont {Higashikawa},\ and\ \citenamefont
  {Ueda}}]{gong2018topological}%
  \BibitemOpen
  \bibfield  {author} {\bibinfo {author} {\bibfnamefont {Z.}~\bibnamefont
  {Gong}}, \bibinfo {author} {\bibfnamefont {Y.}~\bibnamefont {Ashida}},
  \bibinfo {author} {\bibfnamefont {K.}~\bibnamefont {Kawabata}}, \bibinfo
  {author} {\bibfnamefont {K.}~\bibnamefont {Takasan}}, \bibinfo {author}
  {\bibfnamefont {S.}~\bibnamefont {Higashikawa}}, \ and\ \bibinfo {author}
  {\bibfnamefont {M.}~\bibnamefont {Ueda}},\ }\href {\doibase
  10.1103/PhysRevX.8.031079} {\bibfield  {journal} {\bibinfo  {journal}
  {Physical Review X}\ }\textbf {\bibinfo {volume} {8}},\ \bibinfo {pages}
  {031079} (\bibinfo {year} {2018})}\BibitemShut {NoStop}%
\bibitem [{\citenamefont {Lieu}(2018{\natexlab{a}})}]{lieu2018topological}%
  \BibitemOpen
  \bibfield  {author} {\bibinfo {author} {\bibfnamefont {S.}~\bibnamefont
  {Lieu}},\ }\href {\doibase 10.1103/PhysRevB.97.045106} {\bibfield  {journal}
  {\bibinfo  {journal} {Physical Review B}\ }\textbf {\bibinfo {volume} {97}},\
  \bibinfo {pages} {045106} (\bibinfo {year} {2018}{\natexlab{a}})}\BibitemShut
  {NoStop}%
\bibitem [{\citenamefont {Lieu}(2018{\natexlab{b}})}]{lieu2018topological1}%
  \BibitemOpen
  \bibfield  {author} {\bibinfo {author} {\bibfnamefont {S.}~\bibnamefont
  {Lieu}},\ }\href {\doibase 10.1103/PhysRevB.98.115135} {\bibfield  {journal}
  {\bibinfo  {journal} {Physical Review B}\ }\textbf {\bibinfo {volume} {98}},\
  \bibinfo {pages} {115135} (\bibinfo {year} {2018}{\natexlab{b}})}\BibitemShut
  {NoStop}%
\bibitem [{\citenamefont {Kawabata}\ \emph
  {et~al.}(2018{\natexlab{a}})\citenamefont {Kawabata}, \citenamefont
  {Shiozaki},\ and\ \citenamefont {Ueda}}]{kawabata2018anomalous}%
  \BibitemOpen
  \bibfield  {author} {\bibinfo {author} {\bibfnamefont {K.}~\bibnamefont
  {Kawabata}}, \bibinfo {author} {\bibfnamefont {K.}~\bibnamefont {Shiozaki}},
  \ and\ \bibinfo {author} {\bibfnamefont {M.}~\bibnamefont {Ueda}},\ }\href
  {\doibase 10.1103/PhysRevB.98.165148} {\bibfield  {journal} {\bibinfo
  {journal} {Physical Review B}\ }\textbf {\bibinfo {volume} {98}},\ \bibinfo
  {pages} {165148} (\bibinfo {year} {2018}{\natexlab{a}})}\BibitemShut
  {NoStop}%
\bibitem [{\citenamefont {Kawabata}\ \emph
  {et~al.}(2019{\natexlab{a}})\citenamefont {Kawabata}, \citenamefont
  {Higashikawa}, \citenamefont {Gong}, \citenamefont {Ashida},\ and\
  \citenamefont {Ueda}}]{kawabata2019topological}%
  \BibitemOpen
  \bibfield  {author} {\bibinfo {author} {\bibfnamefont {K.}~\bibnamefont
  {Kawabata}}, \bibinfo {author} {\bibfnamefont {S.}~\bibnamefont
  {Higashikawa}}, \bibinfo {author} {\bibfnamefont {Z.}~\bibnamefont {Gong}},
  \bibinfo {author} {\bibfnamefont {Y.}~\bibnamefont {Ashida}}, \ and\ \bibinfo
  {author} {\bibfnamefont {M.}~\bibnamefont {Ueda}},\ }\href {\doibase
  10.1038/s41467-018-08254-y} {\bibfield  {journal} {\bibinfo  {journal}
  {Nature Communications}\ }\textbf {\bibinfo {volume} {10}},\ \bibinfo {pages}
  {297} (\bibinfo {year} {2019}{\natexlab{a}})}\BibitemShut {NoStop}%
\bibitem [{\citenamefont {Kawabata}\ \emph
  {et~al.}(2019{\natexlab{b}})\citenamefont {Kawabata}, \citenamefont
  {Bessho},\ and\ \citenamefont {Sato}}]{kawabata2019non}%
  \BibitemOpen
  \bibfield  {author} {\bibinfo {author} {\bibfnamefont {K.}~\bibnamefont
  {Kawabata}}, \bibinfo {author} {\bibfnamefont {T.}~\bibnamefont {Bessho}}, \
  and\ \bibinfo {author} {\bibfnamefont {M.}~\bibnamefont {Sato}},\ }\href
  {http://arxiv.org/abs/1902.08479 https://arxiv.org/abs/1902.08479} {\bibfield
   {journal} {\bibinfo  {journal} {arXiv preprint arXiv:1902.08479}\ }
  (\bibinfo {year} {2019}{\natexlab{b}})}\BibitemShut {NoStop}%
\bibitem [{\citenamefont {Carlstr{\"{o}}m}\ and\ \citenamefont
  {Bergholtz}(2018)}]{carlstrom2018exceptional}%
  \BibitemOpen
  \bibfield  {author} {\bibinfo {author} {\bibfnamefont {J.}~\bibnamefont
  {Carlstr{\"{o}}m}}\ and\ \bibinfo {author} {\bibfnamefont {E.~J.}\
  \bibnamefont {Bergholtz}},\ }\href {\doibase 10.1103/PhysRevA.98.042114}
  {\bibfield  {journal} {\bibinfo  {journal} {Physical Review A}\ }\textbf
  {\bibinfo {volume} {98}},\ \bibinfo {pages} {042114} (\bibinfo {year}
  {2018})}\BibitemShut {NoStop}%
\bibitem [{\citenamefont {Takata}\ and\ \citenamefont
  {Notomi}(2018)}]{takata2018photonic}%
  \BibitemOpen
  \bibfield  {author} {\bibinfo {author} {\bibfnamefont {K.}~\bibnamefont
  {Takata}}\ and\ \bibinfo {author} {\bibfnamefont {M.}~\bibnamefont
  {Notomi}},\ }\href {\doibase 10.1103/PhysRevLett.121.213902} {\bibfield
  {journal} {\bibinfo  {journal} {Physical Review Letters}\ }\textbf {\bibinfo
  {volume} {121}},\ \bibinfo {pages} {213902} (\bibinfo {year}
  {2018})}\BibitemShut {NoStop}%
\bibitem [{\citenamefont {Moors}\ \emph {et~al.}(2019)\citenamefont {Moors},
  \citenamefont {Zyuzin}, \citenamefont {Zyuzin}, \citenamefont {Tiwari},\ and\
  \citenamefont {Schmidt}}]{moors2019disorder}%
  \BibitemOpen
  \bibfield  {author} {\bibinfo {author} {\bibfnamefont {K.}~\bibnamefont
  {Moors}}, \bibinfo {author} {\bibfnamefont {A.~A.}\ \bibnamefont {Zyuzin}},
  \bibinfo {author} {\bibfnamefont {A.~Y.}\ \bibnamefont {Zyuzin}}, \bibinfo
  {author} {\bibfnamefont {R.~P.}\ \bibnamefont {Tiwari}}, \ and\ \bibinfo
  {author} {\bibfnamefont {T.~L.}\ \bibnamefont {Schmidt}},\ }\href {\doibase
  10.1103/PhysRevB.99.041116} {\bibfield  {journal} {\bibinfo  {journal}
  {Physical Review B}\ }\textbf {\bibinfo {volume} {99}},\ \bibinfo {pages}
  {041116} (\bibinfo {year} {2019})}\BibitemShut {NoStop}%
\bibitem [{\citenamefont {Zyuzin}\ and\ \citenamefont
  {Zyuzin}(2018)}]{zyuzin2018flat}%
  \BibitemOpen
  \bibfield  {author} {\bibinfo {author} {\bibfnamefont {A.~A.}\ \bibnamefont
  {Zyuzin}}\ and\ \bibinfo {author} {\bibfnamefont {A.~Y.}\ \bibnamefont
  {Zyuzin}},\ }\href {\doibase 10.1103/PhysRevB.97.041203} {\bibfield
  {journal} {\bibinfo  {journal} {Physical Review B}\ }\textbf {\bibinfo
  {volume} {97}},\ \bibinfo {pages} {041203} (\bibinfo {year}
  {2018})}\BibitemShut {NoStop}%
\bibitem [{\citenamefont {Kunst}\ and\ \citenamefont
  {Dwivedi}(2019)}]{kunst2019non}%
  \BibitemOpen
  \bibfield  {author} {\bibinfo {author} {\bibfnamefont {F.~K.}\ \bibnamefont
  {Kunst}}\ and\ \bibinfo {author} {\bibfnamefont {V.}~\bibnamefont
  {Dwivedi}},\ }\href {\doibase 10.1103/PhysRevB.99.245116} {\bibfield
  {journal} {\bibinfo  {journal} {Physical Review B}\ }\textbf {\bibinfo
  {volume} {99}},\ \bibinfo {pages} {245116} (\bibinfo {year}
  {2019})}\BibitemShut {NoStop}%
\bibitem [{\citenamefont {Yang}\ and\ \citenamefont {Hu}(2019)}]{yang2019non}%
  \BibitemOpen
  \bibfield  {author} {\bibinfo {author} {\bibfnamefont {Z.}~\bibnamefont
  {Yang}}\ and\ \bibinfo {author} {\bibfnamefont {J.}~\bibnamefont {Hu}},\
  }\href {\doibase 10.1103/PhysRevB.99.081102} {\bibfield  {journal} {\bibinfo
  {journal} {Physical Review B}\ }\textbf {\bibinfo {volume} {99}},\ \bibinfo
  {pages} {081102} (\bibinfo {year} {2019})}\BibitemShut {NoStop}%
\bibitem [{\citenamefont {Zhou}\ \emph {et~al.}(2019)\citenamefont {Zhou},
  \citenamefont {Lee}, \citenamefont {Liu},\ and\ \citenamefont
  {Zhen}}]{zhou2019exceptional}%
  \BibitemOpen
  \bibfield  {author} {\bibinfo {author} {\bibfnamefont {H.}~\bibnamefont
  {Zhou}}, \bibinfo {author} {\bibfnamefont {J.~Y.}\ \bibnamefont {Lee}},
  \bibinfo {author} {\bibfnamefont {S.}~\bibnamefont {Liu}}, \ and\ \bibinfo
  {author} {\bibfnamefont {B.}~\bibnamefont {Zhen}},\ }\href {\doibase
  10.1364/OPTICA.6.000190} {\bibfield  {journal} {\bibinfo  {journal} {Optica}\
  }\textbf {\bibinfo {volume} {6}},\ \bibinfo {pages} {190} (\bibinfo {year}
  {2019})}\BibitemShut {NoStop}%
\bibitem [{\citenamefont {Budich}\ \emph {et~al.}(2019)\citenamefont {Budich},
  \citenamefont {Carlstr{\"{o}}m}, \citenamefont {Kunst},\ and\ \citenamefont
  {Bergholtz}}]{budich2019symmetry}%
  \BibitemOpen
  \bibfield  {author} {\bibinfo {author} {\bibfnamefont {J.~C.}\ \bibnamefont
  {Budich}}, \bibinfo {author} {\bibfnamefont {J.}~\bibnamefont
  {Carlstr{\"{o}}m}}, \bibinfo {author} {\bibfnamefont {F.~K.}\ \bibnamefont
  {Kunst}}, \ and\ \bibinfo {author} {\bibfnamefont {E.~J.}\ \bibnamefont
  {Bergholtz}},\ }\href {\doibase 10.1103/PhysRevB.99.041406} {\bibfield
  {journal} {\bibinfo  {journal} {Physical Review B}\ }\textbf {\bibinfo
  {volume} {99}},\ \bibinfo {pages} {041406} (\bibinfo {year}
  {2019})}\BibitemShut {NoStop}%
\bibitem [{\citenamefont {Okugawa}\ and\ \citenamefont
  {Yokoyama}(2019)}]{okugawa2019topological}%
  \BibitemOpen
  \bibfield  {author} {\bibinfo {author} {\bibfnamefont {R.}~\bibnamefont
  {Okugawa}}\ and\ \bibinfo {author} {\bibfnamefont {T.}~\bibnamefont
  {Yokoyama}},\ }\href {\doibase 10.1103/PhysRevB.99.041202} {\bibfield
  {journal} {\bibinfo  {journal} {Physical Review B}\ }\textbf {\bibinfo
  {volume} {99}},\ \bibinfo {pages} {041202} (\bibinfo {year}
  {2019})}\BibitemShut {NoStop}%
\bibitem [{\citenamefont {Dembowski}\ \emph {et~al.}(2001)\citenamefont
  {Dembowski}, \citenamefont {Gr{\"{a}}f}, \citenamefont {Harney},
  \citenamefont {Heine}, \citenamefont {Heiss}, \citenamefont {Rehfeld},\ and\
  \citenamefont {Richter}}]{dembowski2001experimental}%
  \BibitemOpen
  \bibfield  {author} {\bibinfo {author} {\bibfnamefont {C.}~\bibnamefont
  {Dembowski}}, \bibinfo {author} {\bibfnamefont {H.-D.}\ \bibnamefont
  {Gr{\"{a}}f}}, \bibinfo {author} {\bibfnamefont {H.~L.}\ \bibnamefont
  {Harney}}, \bibinfo {author} {\bibfnamefont {A.}~\bibnamefont {Heine}},
  \bibinfo {author} {\bibfnamefont {W.~D.}\ \bibnamefont {Heiss}}, \bibinfo
  {author} {\bibfnamefont {H.}~\bibnamefont {Rehfeld}}, \ and\ \bibinfo
  {author} {\bibfnamefont {A.}~\bibnamefont {Richter}},\ }\href {\doibase
  10.1103/PhysRevLett.86.787} {\bibfield  {journal} {\bibinfo  {journal}
  {Physical Review Letters}\ }\textbf {\bibinfo {volume} {86}},\ \bibinfo
  {pages} {787} (\bibinfo {year} {2001})}\BibitemShut {NoStop}%
\bibitem [{\citenamefont {Poli}\ \emph {et~al.}(2015)\citenamefont {Poli},
  \citenamefont {Bellec}, \citenamefont {Kuhl}, \citenamefont {Mortessagne},\
  and\ \citenamefont {Schomerus}}]{poli2015selective}%
  \BibitemOpen
  \bibfield  {author} {\bibinfo {author} {\bibfnamefont {C.}~\bibnamefont
  {Poli}}, \bibinfo {author} {\bibfnamefont {M.}~\bibnamefont {Bellec}},
  \bibinfo {author} {\bibfnamefont {U.}~\bibnamefont {Kuhl}}, \bibinfo {author}
  {\bibfnamefont {F.}~\bibnamefont {Mortessagne}}, \ and\ \bibinfo {author}
  {\bibfnamefont {H.}~\bibnamefont {Schomerus}},\ }\href {\doibase
  10.1038/ncomms7710} {\bibfield  {journal} {\bibinfo  {journal} {Nature
  Communications}\ }\textbf {\bibinfo {volume} {6}},\ \bibinfo {pages} {6710}
  (\bibinfo {year} {2015})}\BibitemShut {NoStop}%
\bibitem [{\citenamefont {Zeuner}\ \emph {et~al.}(2015)\citenamefont {Zeuner},
  \citenamefont {Rechtsman}, \citenamefont {Plotnik}, \citenamefont {Lumer},
  \citenamefont {Nolte}, \citenamefont {Rudner}, \citenamefont {Segev},\ and\
  \citenamefont {Szameit}}]{zeuner2015observation}%
  \BibitemOpen
  \bibfield  {author} {\bibinfo {author} {\bibfnamefont {J.~M.}\ \bibnamefont
  {Zeuner}}, \bibinfo {author} {\bibfnamefont {M.~C.}\ \bibnamefont
  {Rechtsman}}, \bibinfo {author} {\bibfnamefont {Y.}~\bibnamefont {Plotnik}},
  \bibinfo {author} {\bibfnamefont {Y.}~\bibnamefont {Lumer}}, \bibinfo
  {author} {\bibfnamefont {S.}~\bibnamefont {Nolte}}, \bibinfo {author}
  {\bibfnamefont {M.~S.}\ \bibnamefont {Rudner}}, \bibinfo {author}
  {\bibfnamefont {M.}~\bibnamefont {Segev}}, \ and\ \bibinfo {author}
  {\bibfnamefont {A.}~\bibnamefont {Szameit}},\ }\href {\doibase
  10.1103/PhysRevLett.115.040402} {\bibfield  {journal} {\bibinfo  {journal}
  {Physical Review Letters}\ }\textbf {\bibinfo {volume} {115}},\ \bibinfo
  {pages} {040402} (\bibinfo {year} {2015})}\BibitemShut {NoStop}%
\bibitem [{\citenamefont {Weimann}\ \emph {et~al.}(2016)\citenamefont
  {Weimann}, \citenamefont {Kremer}, \citenamefont {Plotnik}, \citenamefont
  {Lumer}, \citenamefont {Nolte}, \citenamefont {Makris}, \citenamefont
  {Segev}, \citenamefont {Rechtsman},\ and\ \citenamefont
  {Szameit}}]{weimann2016topologically}%
  \BibitemOpen
  \bibfield  {author} {\bibinfo {author} {\bibfnamefont {S.}~\bibnamefont
  {Weimann}}, \bibinfo {author} {\bibfnamefont {M.}~\bibnamefont {Kremer}},
  \bibinfo {author} {\bibfnamefont {Y.}~\bibnamefont {Plotnik}}, \bibinfo
  {author} {\bibfnamefont {Y.}~\bibnamefont {Lumer}}, \bibinfo {author}
  {\bibfnamefont {S.}~\bibnamefont {Nolte}}, \bibinfo {author} {\bibfnamefont
  {K.~G.}\ \bibnamefont {Makris}}, \bibinfo {author} {\bibfnamefont
  {M.}~\bibnamefont {Segev}}, \bibinfo {author} {\bibfnamefont
  {M.}~\bibnamefont {Rechtsman}}, \ and\ \bibinfo {author} {\bibfnamefont
  {A.}~\bibnamefont {Szameit}},\ }\href {\doibase 10.1038/nmat4811} {\bibfield
  {journal} {\bibinfo  {journal} {Nature Materials}\ }\textbf {\bibinfo
  {volume} {16}},\ \bibinfo {pages} {433} (\bibinfo {year} {2016})}\BibitemShut
  {NoStop}%
\bibitem [{\citenamefont {Zhou}\ \emph {et~al.}(2018)\citenamefont {Zhou},
  \citenamefont {Peng}, \citenamefont {Yoon}, \citenamefont {Hsu},
  \citenamefont {Nelson}, \citenamefont {Fu}, \citenamefont {Joannopoulos},
  \citenamefont {Solja{\v{c}}i{\'{c}}},\ and\ \citenamefont
  {Zhen}}]{zhou2018observation}%
  \BibitemOpen
  \bibfield  {author} {\bibinfo {author} {\bibfnamefont {H.}~\bibnamefont
  {Zhou}}, \bibinfo {author} {\bibfnamefont {C.}~\bibnamefont {Peng}}, \bibinfo
  {author} {\bibfnamefont {Y.}~\bibnamefont {Yoon}}, \bibinfo {author}
  {\bibfnamefont {C.~W.}\ \bibnamefont {Hsu}}, \bibinfo {author} {\bibfnamefont
  {K.~A.}\ \bibnamefont {Nelson}}, \bibinfo {author} {\bibfnamefont
  {L.}~\bibnamefont {Fu}}, \bibinfo {author} {\bibfnamefont {J.~D.}\
  \bibnamefont {Joannopoulos}}, \bibinfo {author} {\bibfnamefont
  {M.}~\bibnamefont {Solja{\v{c}}i{\'{c}}}}, \ and\ \bibinfo {author}
  {\bibfnamefont {B.}~\bibnamefont {Zhen}},\ }\href {\doibase
  10.1126/science.aap9859} {\bibfield  {journal} {\bibinfo  {journal}
  {Science}\ }\textbf {\bibinfo {volume} {359}},\ \bibinfo {pages} {1009}
  (\bibinfo {year} {2018})}\BibitemShut {NoStop}%
\bibitem [{\citenamefont {Cerjan}\ \emph {et~al.}(2019)\citenamefont {Cerjan},
  \citenamefont {Huang}, \citenamefont {Wang}, \citenamefont {Chen},
  \citenamefont {Chong},\ and\ \citenamefont
  {Rechtsman}}]{cerjan2019experimental}%
  \BibitemOpen
  \bibfield  {author} {\bibinfo {author} {\bibfnamefont {A.}~\bibnamefont
  {Cerjan}}, \bibinfo {author} {\bibfnamefont {S.}~\bibnamefont {Huang}},
  \bibinfo {author} {\bibfnamefont {M.}~\bibnamefont {Wang}}, \bibinfo {author}
  {\bibfnamefont {K.~P.}\ \bibnamefont {Chen}}, \bibinfo {author}
  {\bibfnamefont {Y.}~\bibnamefont {Chong}}, \ and\ \bibinfo {author}
  {\bibfnamefont {M.~C.}\ \bibnamefont {Rechtsman}},\ }\href {\doibase
  10.1038/s41566-019-0453-z} {\bibfield  {journal} {\bibinfo  {journal} {Nature
  Photonics}\ } (\bibinfo {year} {2019}),\
  10.1038/s41566-019-0453-z}\BibitemShut {NoStop}%
\bibitem [{\citenamefont {Xiong}(2018)}]{xiong2018why}%
  \BibitemOpen
  \bibfield  {author} {\bibinfo {author} {\bibfnamefont {Y.}~\bibnamefont
  {Xiong}},\ }\href {\doibase 10.1088/2399-6528/aab64a} {\bibfield  {journal}
  {\bibinfo  {journal} {Journal of Physics Communications}\ }\textbf {\bibinfo
  {volume} {2}},\ \bibinfo {pages} {035043} (\bibinfo {year}
  {2018})}\BibitemShut {NoStop}%
\bibitem [{\citenamefont {Kunst}\ \emph {et~al.}(2018)\citenamefont {Kunst},
  \citenamefont {Edvardsson}, \citenamefont {Budich},\ and\ \citenamefont
  {Bergholtz}}]{kunst2018biorthogonal}%
  \BibitemOpen
  \bibfield  {author} {\bibinfo {author} {\bibfnamefont {F.~K.}\ \bibnamefont
  {Kunst}}, \bibinfo {author} {\bibfnamefont {E.}~\bibnamefont {Edvardsson}},
  \bibinfo {author} {\bibfnamefont {J.~C.}\ \bibnamefont {Budich}}, \ and\
  \bibinfo {author} {\bibfnamefont {E.~J.}\ \bibnamefont {Bergholtz}},\ }\href
  {\doibase 10.1103/PhysRevLett.121.026808} {\bibfield  {journal} {\bibinfo
  {journal} {Physical Review Letters}\ }\textbf {\bibinfo {volume} {121}},\
  \bibinfo {pages} {026808} (\bibinfo {year} {2018})}\BibitemShut {NoStop}%
\bibitem [{\citenamefont {Yao}\ and\ \citenamefont {Wang}(2018)}]{yao2018edge}%
  \BibitemOpen
  \bibfield  {author} {\bibinfo {author} {\bibfnamefont {S.}~\bibnamefont
  {Yao}}\ and\ \bibinfo {author} {\bibfnamefont {Z.}~\bibnamefont {Wang}},\
  }\href {\doibase 10.1103/PhysRevLett.121.086803} {\bibfield  {journal}
  {\bibinfo  {journal} {Physical Review Letters}\ }\textbf {\bibinfo {volume}
  {121}},\ \bibinfo {pages} {086803} (\bibinfo {year} {2018})}\BibitemShut
  {NoStop}%
\bibitem [{\citenamefont {Yao}\ \emph {et~al.}(2018)\citenamefont {Yao},
  \citenamefont {Song},\ and\ \citenamefont {Wang}}]{yao2018non}%
  \BibitemOpen
  \bibfield  {author} {\bibinfo {author} {\bibfnamefont {S.}~\bibnamefont
  {Yao}}, \bibinfo {author} {\bibfnamefont {F.}~\bibnamefont {Song}}, \ and\
  \bibinfo {author} {\bibfnamefont {Z.}~\bibnamefont {Wang}},\ }\href {\doibase
  10.1103/PhysRevLett.121.136802} {\bibfield  {journal} {\bibinfo  {journal}
  {Physical Review Letters}\ }\textbf {\bibinfo {volume} {121}},\ \bibinfo
  {pages} {136802} (\bibinfo {year} {2018})}\BibitemShut {NoStop}%
\bibitem [{\citenamefont {Lee}\ and\ \citenamefont
  {Thomale}(2019)}]{lee2019anatomy}%
  \BibitemOpen
  \bibfield  {author} {\bibinfo {author} {\bibfnamefont {C.~H.}\ \bibnamefont
  {Lee}}\ and\ \bibinfo {author} {\bibfnamefont {R.}~\bibnamefont {Thomale}},\
  }\href {\doibase 10.1103/PhysRevB.99.201103} {\bibfield  {journal} {\bibinfo
  {journal} {Physical Review B}\ }\textbf {\bibinfo {volume} {99}},\ \bibinfo
  {pages} {201103} (\bibinfo {year} {2019})}\BibitemShut {NoStop}%
\bibitem [{\citenamefont {Borgnia}\ \emph {et~al.}(2019)\citenamefont
  {Borgnia}, \citenamefont {Kruchkov},\ and\ \citenamefont
  {Slager}}]{borgnia2019non}%
  \BibitemOpen
  \bibfield  {author} {\bibinfo {author} {\bibfnamefont {D.~S.}\ \bibnamefont
  {Borgnia}}, \bibinfo {author} {\bibfnamefont {A.~J.}\ \bibnamefont
  {Kruchkov}}, \ and\ \bibinfo {author} {\bibfnamefont {R.-J.}\ \bibnamefont
  {Slager}},\ }\href {http://arxiv.org/abs/1902.07217} {\bibfield  {journal}
  {\bibinfo  {journal} {arXiv preprint arXiv:1902.07217}\ } (\bibinfo {year}
  {2019})}\BibitemShut {NoStop}%
\bibitem [{\citenamefont {Zirnstein}\ \emph {et~al.}(2019)\citenamefont
  {Zirnstein}, \citenamefont {Refael},\ and\ \citenamefont
  {Rosenow}}]{zirnstein2019bulk}%
  \BibitemOpen
  \bibfield  {author} {\bibinfo {author} {\bibfnamefont {H.-G.}\ \bibnamefont
  {Zirnstein}}, \bibinfo {author} {\bibfnamefont {G.}~\bibnamefont {Refael}}, \
  and\ \bibinfo {author} {\bibfnamefont {B.}~\bibnamefont {Rosenow}},\ }\href
  {http://arxiv.org/abs/1901.11241} {\bibfield  {journal} {\bibinfo  {journal}
  {arXiv preprint arXiv:1901.11241}\ } (\bibinfo {year} {2019})}\BibitemShut
  {NoStop}%
\bibitem [{\citenamefont {Herviou}\ \emph {et~al.}(2019)\citenamefont
  {Herviou}, \citenamefont {Bardarson},\ and\ \citenamefont
  {Regnault}}]{herviou2019defining}%
  \BibitemOpen
  \bibfield  {author} {\bibinfo {author} {\bibfnamefont {L.}~\bibnamefont
  {Herviou}}, \bibinfo {author} {\bibfnamefont {J.~H.}\ \bibnamefont
  {Bardarson}}, \ and\ \bibinfo {author} {\bibfnamefont {N.}~\bibnamefont
  {Regnault}},\ }\href {\doibase 10.1103/PhysRevA.99.052118} {\bibfield
  {journal} {\bibinfo  {journal} {Physical Review A}\ }\textbf {\bibinfo
  {volume} {99}},\ \bibinfo {pages} {052118} (\bibinfo {year}
  {2019})}\BibitemShut {NoStop}%
\bibitem [{\citenamefont {Ge}\ \emph {et~al.}(2019)\citenamefont {Ge},
  \citenamefont {Zhang}, \citenamefont {Liu}, \citenamefont {Li}, \citenamefont
  {Fan},\ and\ \citenamefont {Nori}}]{ge2019topological}%
  \BibitemOpen
  \bibfield  {author} {\bibinfo {author} {\bibfnamefont {Z.-Y.}\ \bibnamefont
  {Ge}}, \bibinfo {author} {\bibfnamefont {Y.-R.}\ \bibnamefont {Zhang}},
  \bibinfo {author} {\bibfnamefont {T.}~\bibnamefont {Liu}}, \bibinfo {author}
  {\bibfnamefont {S.-W.}\ \bibnamefont {Li}}, \bibinfo {author} {\bibfnamefont
  {H.}~\bibnamefont {Fan}}, \ and\ \bibinfo {author} {\bibfnamefont
  {F.}~\bibnamefont {Nori}},\ }\href {http://arxiv.org/abs/1903.09985}
  {\bibfield  {journal} {\bibinfo  {journal} {arXiv preprint arXiv:1903.09985}\
  } (\bibinfo {year} {2019})}\BibitemShut {NoStop}%
\bibitem [{\citenamefont {Liu}\ \emph {et~al.}(2019)\citenamefont {Liu},
  \citenamefont {Zhang}, \citenamefont {Ai}, \citenamefont {Gong},
  \citenamefont {Kawabata}, \citenamefont {Ueda},\ and\ \citenamefont
  {Nori}}]{liu2019second}%
  \BibitemOpen
  \bibfield  {author} {\bibinfo {author} {\bibfnamefont {T.}~\bibnamefont
  {Liu}}, \bibinfo {author} {\bibfnamefont {Y.~R.}\ \bibnamefont {Zhang}},
  \bibinfo {author} {\bibfnamefont {Q.}~\bibnamefont {Ai}}, \bibinfo {author}
  {\bibfnamefont {Z.}~\bibnamefont {Gong}}, \bibinfo {author} {\bibfnamefont
  {K.}~\bibnamefont {Kawabata}}, \bibinfo {author} {\bibfnamefont
  {M.}~\bibnamefont {Ueda}}, \ and\ \bibinfo {author} {\bibfnamefont
  {F.}~\bibnamefont {Nori}},\ }\href {\doibase 10.1103/PhysRevLett.122.076801}
  {\bibfield  {journal} {\bibinfo  {journal} {Physical Review Letters}\
  }\textbf {\bibinfo {volume} {122}},\ \bibinfo {pages} {076801} (\bibinfo
  {year} {2019})}\BibitemShut {NoStop}%
\bibitem [{\citenamefont {{Longhi}}(2019)}]{longhi2019}%
  \BibitemOpen
  \bibfield  {author} {\bibinfo {author} {\bibfnamefont {S.}~\bibnamefont
  {{Longhi}}},\ }\href@noop {} {\bibfield  {journal} {\bibinfo  {journal}
  {arXiv preprint arXiv:1905.09460}\ } (\bibinfo {year} {2019})}\BibitemShut
  {NoStop}%
\bibitem [{\citenamefont {{Wu}}\ and\ \citenamefont
  {{Hou}}(2019)}]{WuDefect2019}%
  \BibitemOpen
  \bibfield  {author} {\bibinfo {author} {\bibfnamefont {Y.-J.}\ \bibnamefont
  {{Wu}}}\ and\ \bibinfo {author} {\bibfnamefont {J.}~\bibnamefont {{Hou}}},\
  }\href@noop {} {\bibfield  {journal} {\bibinfo  {journal} {arXiv preprint
  arXiv:1905.09346}\ } (\bibinfo {year} {2019})}\BibitemShut {NoStop}%
\bibitem [{\citenamefont {{Yuce}}(2019)}]{YuceSkin2019}%
  \BibitemOpen
  \bibfield  {author} {\bibinfo {author} {\bibfnamefont {C.}~\bibnamefont
  {{Yuce}}},\ }\href@noop {} {\bibfield  {journal} {\bibinfo  {journal} {arXiv
  preprint arXiv:1905.09328}\ } (\bibinfo {year} {2019})}\BibitemShut {NoStop}%
\bibitem [{\citenamefont {{Yamamoto}}\ \emph {et~al.}(2019)\citenamefont
  {{Yamamoto}}, \citenamefont {{Nakagawa}}, \citenamefont {{Adachi}},
  \citenamefont {{Takasan}}, \citenamefont {{Ueda}},\ and\ \citenamefont
  {{Kawakami}}}]{YamamotoNHSF2019}%
  \BibitemOpen
  \bibfield  {author} {\bibinfo {author} {\bibfnamefont {K.}~\bibnamefont
  {{Yamamoto}}}, \bibinfo {author} {\bibfnamefont {M.}~\bibnamefont
  {{Nakagawa}}}, \bibinfo {author} {\bibfnamefont {K.}~\bibnamefont
  {{Adachi}}}, \bibinfo {author} {\bibfnamefont {K.}~\bibnamefont {{Takasan}}},
  \bibinfo {author} {\bibfnamefont {M.}~\bibnamefont {{Ueda}}}, \ and\ \bibinfo
  {author} {\bibfnamefont {N.}~\bibnamefont {{Kawakami}}},\ }\href@noop {}
  {\bibfield  {journal} {\bibinfo  {journal} {arXiv preprint arXiv:1903.04720}\
  } (\bibinfo {year} {2019})}\BibitemShut {NoStop}%
\bibitem [{\citenamefont {{Deng}}\ and\ \citenamefont
  {{Yi}}(2019)}]{DengSSH2019}%
  \BibitemOpen
  \bibfield  {author} {\bibinfo {author} {\bibfnamefont {T.-S.}\ \bibnamefont
  {{Deng}}}\ and\ \bibinfo {author} {\bibfnamefont {W.}~\bibnamefont {{Yi}}},\
  }\href@noop {} {\bibfield  {journal} {\bibinfo  {journal} {arXiv preprint
  arXiv:1903.03811}\ } (\bibinfo {year} {2019})}\BibitemShut {NoStop}%
\bibitem [{\citenamefont {Bliokh}\ \emph {et~al.}(2019)\citenamefont {Bliokh},
  \citenamefont {Leykam}, \citenamefont {Lein},\ and\ \citenamefont
  {Nori}}]{Bliokh2019topological}%
  \BibitemOpen
  \bibfield  {author} {\bibinfo {author} {\bibfnamefont {K.~Y.}\ \bibnamefont
  {Bliokh}}, \bibinfo {author} {\bibfnamefont {D.}~\bibnamefont {Leykam}},
  \bibinfo {author} {\bibfnamefont {M.}~\bibnamefont {Lein}}, \ and\ \bibinfo
  {author} {\bibfnamefont {F.}~\bibnamefont {Nori}},\ }\href {\doibase
  10.1038/s41467-019-08397-6} {\bibfield  {journal} {\bibinfo  {journal}
  {Nature Communications}\ }\textbf {\bibinfo {volume} {10}},\ \bibinfo {pages}
  {580} (\bibinfo {year} {2019})}\BibitemShut {NoStop}%
\bibitem [{\citenamefont {Song}\ \emph {et~al.}(2019)\citenamefont {Song},
  \citenamefont {Yao},\ and\ \citenamefont {Wang}}]{song2019non}%
  \BibitemOpen
  \bibfield  {author} {\bibinfo {author} {\bibfnamefont {F.}~\bibnamefont
  {Song}}, \bibinfo {author} {\bibfnamefont {S.}~\bibnamefont {Yao}}, \ and\
  \bibinfo {author} {\bibfnamefont {Z.}~\bibnamefont {Wang}},\ }\href
  {http://arxiv.org/abs/1904.08432} {\bibfield  {journal} {\bibinfo  {journal}
  {arXiv preprint arXiv:1904.08432}\ } (\bibinfo {year} {2019})}\BibitemShut
  {NoStop}%
\bibitem [{\citenamefont {El-Ganainy}\ \emph {et~al.}(2018)\citenamefont
  {El-Ganainy}, \citenamefont {Makris}, \citenamefont {Khajavikhan},
  \citenamefont {Musslimani}, \citenamefont {Rotter},\ and\ \citenamefont
  {Christodoulides}}]{el-ganainy2018non}%
  \BibitemOpen
  \bibfield  {author} {\bibinfo {author} {\bibfnamefont {R.}~\bibnamefont
  {El-Ganainy}}, \bibinfo {author} {\bibfnamefont {K.~G.}\ \bibnamefont
  {Makris}}, \bibinfo {author} {\bibfnamefont {M.}~\bibnamefont {Khajavikhan}},
  \bibinfo {author} {\bibfnamefont {Z.~H.}\ \bibnamefont {Musslimani}},
  \bibinfo {author} {\bibfnamefont {S.}~\bibnamefont {Rotter}}, \ and\ \bibinfo
  {author} {\bibfnamefont {D.~N.}\ \bibnamefont {Christodoulides}},\ }\href
  {\doibase 10.1038/nphys4323} {\bibfield  {journal} {\bibinfo  {journal}
  {Nature Physics}\ }\textbf {\bibinfo {volume} {14}},\ \bibinfo {pages} {11}
  (\bibinfo {year} {2018})}\BibitemShut {NoStop}%
\bibitem [{\citenamefont {Konotop}\ \emph {et~al.}(2016)\citenamefont
  {Konotop}, \citenamefont {Yang},\ and\ \citenamefont
  {Zezyulin}}]{konotop2016nonlinear}%
  \BibitemOpen
  \bibfield  {author} {\bibinfo {author} {\bibfnamefont {V.~V.}\ \bibnamefont
  {Konotop}}, \bibinfo {author} {\bibfnamefont {J.}~\bibnamefont {Yang}}, \
  and\ \bibinfo {author} {\bibfnamefont {D.~A.}\ \bibnamefont {Zezyulin}},\
  }\href {\doibase 10.1103/RevModPhys.88.035002} {\bibfield  {journal}
  {\bibinfo  {journal} {Reviews of Modern Physics}\ }\textbf {\bibinfo {volume}
  {88}},\ \bibinfo {pages} {035002} (\bibinfo {year} {2016})}\BibitemShut
  {NoStop}%
\bibitem [{\citenamefont {Cao}\ and\ \citenamefont
  {Wiersig}(2015)}]{cao2015dielectric}%
  \BibitemOpen
  \bibfield  {author} {\bibinfo {author} {\bibfnamefont {H.}~\bibnamefont
  {Cao}}\ and\ \bibinfo {author} {\bibfnamefont {J.}~\bibnamefont {Wiersig}},\
  }\href {\doibase 10.1103/RevModPhys.87.61} {\bibfield  {journal} {\bibinfo
  {journal} {Reviews of Modern Physics}\ }\textbf {\bibinfo {volume} {87}},\
  \bibinfo {pages} {61} (\bibinfo {year} {2015})}\BibitemShut {NoStop}%
\bibitem [{\citenamefont {Doppler}\ \emph {et~al.}(2016)\citenamefont
  {Doppler}, \citenamefont {Mailybaev}, \citenamefont {B{\"{o}}hm},
  \citenamefont {Kuhl}, \citenamefont {Girschik}, \citenamefont {Libisch},
  \citenamefont {Milburn}, \citenamefont {Rabl}, \citenamefont {Moiseyev},\
  and\ \citenamefont {Rotter}}]{doppler2016dynamically}%
  \BibitemOpen
  \bibfield  {author} {\bibinfo {author} {\bibfnamefont {J.}~\bibnamefont
  {Doppler}}, \bibinfo {author} {\bibfnamefont {A.~A.}\ \bibnamefont
  {Mailybaev}}, \bibinfo {author} {\bibfnamefont {J.}~\bibnamefont
  {B{\"{o}}hm}}, \bibinfo {author} {\bibfnamefont {U.}~\bibnamefont {Kuhl}},
  \bibinfo {author} {\bibfnamefont {A.}~\bibnamefont {Girschik}}, \bibinfo
  {author} {\bibfnamefont {F.}~\bibnamefont {Libisch}}, \bibinfo {author}
  {\bibfnamefont {T.~J.}\ \bibnamefont {Milburn}}, \bibinfo {author}
  {\bibfnamefont {P.}~\bibnamefont {Rabl}}, \bibinfo {author} {\bibfnamefont
  {N.}~\bibnamefont {Moiseyev}}, \ and\ \bibinfo {author} {\bibfnamefont
  {S.}~\bibnamefont {Rotter}},\ }\href {http://dx.doi.org/10.1038/nature18605
  http://10.0.4.14/nature18605
  http://www.nature.com/nature/journal/v537/n7618/abs/nature18605.html{\#}supplementary-information}
  {\bibfield  {journal} {\bibinfo  {journal} {Nature}\ }\textbf {\bibinfo
  {volume} {537}},\ \bibinfo {pages} {76} (\bibinfo {year} {2016})}\BibitemShut
  {NoStop}%
\bibitem [{\citenamefont {Xu}\ \emph {et~al.}(2016)\citenamefont {Xu},
  \citenamefont {Mason}, \citenamefont {Jiang},\ and\ \citenamefont
  {Harris}}]{xu2016topological}%
  \BibitemOpen
  \bibfield  {author} {\bibinfo {author} {\bibfnamefont {H.}~\bibnamefont
  {Xu}}, \bibinfo {author} {\bibfnamefont {D.}~\bibnamefont {Mason}}, \bibinfo
  {author} {\bibfnamefont {L.}~\bibnamefont {Jiang}}, \ and\ \bibinfo {author}
  {\bibfnamefont {J.~G.~E.}\ \bibnamefont {Harris}},\ }\href
  {http://dx.doi.org/10.1038/nature18604 http://10.0.4.14/nature18604
  http://arxiv.org/abs/1602.06881} {\bibfield  {journal} {\bibinfo  {journal}
  {Nature}\ }\textbf {\bibinfo {volume} {537}},\ \bibinfo {pages} {80}
  (\bibinfo {year} {2016})}\BibitemShut {NoStop}%
\bibitem [{\citenamefont {Lapp}\ \emph {et~al.}(2019)\citenamefont {Lapp},
  \citenamefont {Ang'ong'a}, \citenamefont {An},\ and\ \citenamefont
  {Gadway}}]{lapp2019engineering}%
  \BibitemOpen
  \bibfield  {author} {\bibinfo {author} {\bibfnamefont {S.}~\bibnamefont
  {Lapp}}, \bibinfo {author} {\bibfnamefont {J.}~\bibnamefont {Ang'ong'a}},
  \bibinfo {author} {\bibfnamefont {F.~A.}\ \bibnamefont {An}}, \ and\ \bibinfo
  {author} {\bibfnamefont {B.}~\bibnamefont {Gadway}},\ }\href {\doibase
  10.1088/1367-2630/ab1147} {\bibfield  {journal} {\bibinfo  {journal} {New
  Journal of Physics}\ }\textbf {\bibinfo {volume} {21}},\ \bibinfo {pages}
  {045006} (\bibinfo {year} {2019})}\BibitemShut {NoStop}%
\bibitem [{\citenamefont {Amir}\ \emph {et~al.}(2016)\citenamefont {Amir},
  \citenamefont {Hatano},\ and\ \citenamefont {Nelson}}]{amir2016non}%
  \BibitemOpen
  \bibfield  {author} {\bibinfo {author} {\bibfnamefont {A.}~\bibnamefont
  {Amir}}, \bibinfo {author} {\bibfnamefont {N.}~\bibnamefont {Hatano}}, \ and\
  \bibinfo {author} {\bibfnamefont {D.~R.}\ \bibnamefont {Nelson}},\ }\href
  {\doibase 10.1103/PhysRevE.93.042310} {\bibfield  {journal} {\bibinfo
  {journal} {Physical Review E}\ }\textbf {\bibinfo {volume} {93}},\ \bibinfo
  {pages} {042310} (\bibinfo {year} {2016})}\BibitemShut {NoStop}%
\bibitem [{\citenamefont {McDonald}\ \emph {et~al.}(2018)\citenamefont
  {McDonald}, \citenamefont {Pereg-Barnea},\ and\ \citenamefont
  {Clerk}}]{mcdonald2018phase}%
  \BibitemOpen
  \bibfield  {author} {\bibinfo {author} {\bibfnamefont {A.}~\bibnamefont
  {McDonald}}, \bibinfo {author} {\bibfnamefont {T.}~\bibnamefont
  {Pereg-Barnea}}, \ and\ \bibinfo {author} {\bibfnamefont {A.}~\bibnamefont
  {Clerk}},\ }\href {\doibase 10.1103/PhysRevX.8.041031} {\bibfield  {journal}
  {\bibinfo  {journal} {Physical Review X}\ }\textbf {\bibinfo {volume} {8}},\
  \bibinfo {pages} {041031} (\bibinfo {year} {2018})}\BibitemShut {NoStop}%
\bibitem [{\citenamefont {{Lu}}\ and\ \citenamefont
  {{Lu}}(2018)}]{YML_magnon2018}%
  \BibitemOpen
  \bibfield  {author} {\bibinfo {author} {\bibfnamefont {F.}~\bibnamefont
  {{Lu}}}\ and\ \bibinfo {author} {\bibfnamefont {Y.-M.}\ \bibnamefont
  {{Lu}}},\ }\href@noop {} {\bibfield  {journal} {\bibinfo  {journal} {arXiv
  preprint arXiv:1807.05232}\ } (\bibinfo {year} {2018})}\BibitemShut {NoStop}%
\bibitem [{\citenamefont {Kozii}\ and\ \citenamefont
  {Fu}(2017)}]{kozii2017non}%
  \BibitemOpen
  \bibfield  {author} {\bibinfo {author} {\bibfnamefont {V.}~\bibnamefont
  {Kozii}}\ and\ \bibinfo {author} {\bibfnamefont {L.}~\bibnamefont {Fu}},\
  }\href {https://arxiv.org/abs/1708.05841 http://arxiv.org/abs/1708.05841}
  {\bibfield  {journal} {\bibinfo  {journal} {arXiv preprint arXiv:1708.05841}\
  } (\bibinfo {year} {2017})}\BibitemShut {NoStop}%
\bibitem [{\citenamefont {Zhou}\ and\ \citenamefont
  {Lee}(2019)}]{zhou2019periodic}%
  \BibitemOpen
  \bibfield  {author} {\bibinfo {author} {\bibfnamefont {H.}~\bibnamefont
  {Zhou}}\ and\ \bibinfo {author} {\bibfnamefont {J.~Y.}\ \bibnamefont {Lee}},\
  }\href {\doibase 10.1103/PhysRevB.99.235112} {\bibfield  {journal} {\bibinfo
  {journal} {Physical Review B}\ }\textbf {\bibinfo {volume} {99}},\ \bibinfo
  {pages} {235112} (\bibinfo {year} {2019})}\BibitemShut {NoStop}%
\bibitem [{\citenamefont {Kawabata}\ \emph
  {et~al.}(2018{\natexlab{b}})\citenamefont {Kawabata}, \citenamefont
  {Shiozaki}, \citenamefont {Ueda},\ and\ \citenamefont
  {Sato}}]{kawabata2018symmetry}%
  \BibitemOpen
  \bibfield  {author} {\bibinfo {author} {\bibfnamefont {K.}~\bibnamefont
  {Kawabata}}, \bibinfo {author} {\bibfnamefont {K.}~\bibnamefont {Shiozaki}},
  \bibinfo {author} {\bibfnamefont {M.}~\bibnamefont {Ueda}}, \ and\ \bibinfo
  {author} {\bibfnamefont {M.}~\bibnamefont {Sato}},\ }\href
  {https://arxiv.org/abs/1812.09133 http://arxiv.org/abs/1812.09133} {\bibfield
   {journal} {\bibinfo  {journal} {arXiv preprint arXiv:1812.09133}\ }
  (\bibinfo {year} {2018}{\natexlab{b}})}\BibitemShut {NoStop}%
\bibitem [{\citenamefont {Altland}\ and\ \citenamefont
  {Zirnbauer}(1997)}]{altland1997nonstandard}%
  \BibitemOpen
  \bibfield  {author} {\bibinfo {author} {\bibfnamefont {A.}~\bibnamefont
  {Altland}}\ and\ \bibinfo {author} {\bibfnamefont {M.~R.}\ \bibnamefont
  {Zirnbauer}},\ }\href {\doibase 10.1103/PhysRevB.55.1142} {\bibfield
  {journal} {\bibinfo  {journal} {Physical Review B}\ }\textbf {\bibinfo
  {volume} {55}},\ \bibinfo {pages} {1142} (\bibinfo {year}
  {1997})}\BibitemShut {NoStop}%
\bibitem [{\citenamefont {Ryu}\ \emph {et~al.}(2010)\citenamefont {Ryu},
  \citenamefont {Schnyder}, \citenamefont {Furusaki},\ and\ \citenamefont
  {Ludwig}}]{ryu2010topological}%
  \BibitemOpen
  \bibfield  {author} {\bibinfo {author} {\bibfnamefont {S.}~\bibnamefont
  {Ryu}}, \bibinfo {author} {\bibfnamefont {A.~P.}\ \bibnamefont {Schnyder}},
  \bibinfo {author} {\bibfnamefont {A.}~\bibnamefont {Furusaki}}, \ and\
  \bibinfo {author} {\bibfnamefont {A.~W.~W.}\ \bibnamefont {Ludwig}},\ }\href
  {\doibase 10.1088/1367-2630/12/6/065010} {\bibfield  {journal} {\bibinfo
  {journal} {New Journal of Physics}\ }\textbf {\bibinfo {volume} {12}},\
  \bibinfo {pages} {065010} (\bibinfo {year} {2010})}\BibitemShut {NoStop}%
\bibitem [{\citenamefont {Bernard}\ and\ \citenamefont
  {LeClair}(2002)}]{Bernard2001}%
  \BibitemOpen
  \bibfield  {author} {\bibinfo {author} {\bibfnamefont {D.}~\bibnamefont
  {Bernard}}\ and\ \bibinfo {author} {\bibfnamefont {A.}~\bibnamefont
  {LeClair}},\ }in\ \href {\doibase 10.1007/978-94-010-0514-2_19} {\emph
  {\bibinfo {booktitle} {Statistical Field Theories}}},\ \bibinfo {editor}
  {edited by\ \bibinfo {editor} {\bibfnamefont {A.}~\bibnamefont {Cappelli}}\
  and\ \bibinfo {editor} {\bibfnamefont {G.}~\bibnamefont {Mussardo}}}\
  (\bibinfo  {publisher} {Springer},\ \bibinfo {year} {2002})\ \Eprint
  {http://arxiv.org/abs/0110649} {arXiv:0110649 [cond-mat]} \BibitemShut
  {NoStop}%
\bibitem [{\citenamefont {Sato}\ \emph {et~al.}(2012)\citenamefont {Sato},
  \citenamefont {Hasebe}, \citenamefont {Esaki},\ and\ \citenamefont
  {Kohmoto}}]{sato2012time}%
  \BibitemOpen
  \bibfield  {author} {\bibinfo {author} {\bibfnamefont {M.}~\bibnamefont
  {Sato}}, \bibinfo {author} {\bibfnamefont {K.}~\bibnamefont {Hasebe}},
  \bibinfo {author} {\bibfnamefont {K.}~\bibnamefont {Esaki}}, \ and\ \bibinfo
  {author} {\bibfnamefont {M.}~\bibnamefont {Kohmoto}},\ }\href {\doibase
  10.1143/PTP.127.937} {\bibfield  {journal} {\bibinfo  {journal} {Progress of
  Theoretical Physics}\ }\textbf {\bibinfo {volume} {127}},\ \bibinfo {pages}
  {937} (\bibinfo {year} {2012})}\BibitemShut {NoStop}%
\bibitem [{\citenamefont {{DeMarco}}\ and\ \citenamefont
  {{Wen}}(2018)}]{DeMarco2018}%
  \BibitemOpen
  \bibfield  {author} {\bibinfo {author} {\bibfnamefont {M.}~\bibnamefont
  {{DeMarco}}}\ and\ \bibinfo {author} {\bibfnamefont {X.-G.}\ \bibnamefont
  {{Wen}}},\ }\href@noop {} {\bibfield  {journal} {\bibinfo  {journal} {arXiv
  preprint arXiv:1805.03663}\ } (\bibinfo {year} {2018})}\BibitemShut {NoStop}%
\bibitem [{SM()}]{SM}%
  \BibitemOpen
  \href@noop {} {}\bibinfo {note} {See Supplemental Material}\BibitemShut
  {NoStop}%
\bibitem [{\citenamefont {Teo}\ and\ \citenamefont
  {Kane}(2010)}]{teo2010topological}%
  \BibitemOpen
  \bibfield  {author} {\bibinfo {author} {\bibfnamefont {J.~C.}\ \bibnamefont
  {Teo}}\ and\ \bibinfo {author} {\bibfnamefont {C.~L.}\ \bibnamefont {Kane}},\
  }\href {\doibase 10.1103/PhysRevB.82.115120} {\bibfield  {journal} {\bibinfo
  {journal} {Physical Review B}\ }\textbf {\bibinfo {volume} {82}},\ \bibinfo
  {pages} {115120} (\bibinfo {year} {2010})}\BibitemShut {NoStop}%
\bibitem [{Note1()}]{Note1}%
  \BibitemOpen
  \bibinfo {note} {More precisely, $\theta $ is a latitude for this higher
  dimensional Brillouin zone, which is given by the suspension of the original
  one}\BibitemShut {NoStop}%
\bibitem [{Note2()}]{Note2}%
  \BibitemOpen
  \bibinfo {note} {\label {footnote:deform}Lorentz symmetry is broken in
  crystals, so Dirac and Weyl fermions that require four-fold or more
  degeneracy are not stabilized in higher dimensions. Instead, anomalous
  boundary states (e.g. surface) appear as symmetry-preserving deformations of
  Dirac or Weyl fermions, with the same anomaly as the Dirac or Weyl
  fermions.}\BibitemShut {Stop}%
\bibitem [{Note3()}]{Note3}%
  \BibitemOpen
  \bibinfo {note} {$k \Gamma _i^* + \Gamma _i k = 0$ and $c \Gamma _i^T -
  \Gamma _i c = 0$}\BibitemShut {NoStop}%
\bibitem [{Note4()}]{Note4}%
  \BibitemOpen
  \bibinfo {note} {Let $f_0=\gamma ({\protect \bf k})$ and $f_i=\sin k_i$.
  Then, $\protect \mathaccentV {bar}016{H}: \protect \textrm {BZ}^d \rightarrow
  (f_0,...,f_d)$ is homotopic to a trivial constant map $g = (\gamma _0, 0,
  ...,0)$ with $\gamma _0 < 0$. This can be proven by constructing the
  following homotopy map \begin {equation} F(\protect \mathbf {k},t) = (1-t)
  f(\protect \mathbf {k}) + t g(\protect \mathbf {k}), \end {equation} which
  gives a fully gapped Hamiltonian for $(\protect \mathbf {k},t) \in \protect
  \textrm {BZ}^d \times [0,1]$, where $F(k,0) = f(k)$ and $F(k,1) =
  g(k)$.}\BibitemShut {Stop}%
\bibitem [{\citenamefont {Ashida}\ and\ \citenamefont
  {Ueda}(2018)}]{Ashida2018_jump}%
  \BibitemOpen
  \bibfield  {author} {\bibinfo {author} {\bibfnamefont {Y.}~\bibnamefont
  {Ashida}}\ and\ \bibinfo {author} {\bibfnamefont {M.}~\bibnamefont {Ueda}},\
  }\href {\doibase 10.1103/PhysRevLett.120.185301} {\bibfield  {journal}
  {\bibinfo  {journal} {Phys. Rev. Lett.}\ }\textbf {\bibinfo {volume} {120}},\
  \bibinfo {pages} {185301} (\bibinfo {year} {2018})}\BibitemShut {NoStop}%
\bibitem [{\citenamefont {Jackiw}\ and\ \citenamefont
  {Rebbi}(1976)}]{JackiwRebbi}%
  \BibitemOpen
  \bibfield  {author} {\bibinfo {author} {\bibfnamefont {R.}~\bibnamefont
  {Jackiw}}\ and\ \bibinfo {author} {\bibfnamefont {C.}~\bibnamefont {Rebbi}},\
  }\href {\doibase 10.1103/PhysRevD.13.3398} {\bibfield  {journal} {\bibinfo
  {journal} {Phys. Rev. D}\ }\textbf {\bibinfo {volume} {13}},\ \bibinfo
  {pages} {3398} (\bibinfo {year} {1976})}\BibitemShut {NoStop}%
\bibitem [{Note5()}]{Note5}%
  \BibitemOpen
  \bibinfo {note} {As a physical Hilbert space consists of normalizable
  wavefunctions, we should only take decaying solutions.}\BibitemShut {Stop}%
\end{thebibliography}%

\clearpage
\newpage

\setcounter{equation}{0}
\setcounter{figure}{0}
\setcounter{table}{0}
\setcounter{page}{1}

\makeatletter
\renewcommand{\theequation}{S\arabic{equation}}
\renewcommand{\thefigure}{S\arabic{figure}}
\renewcommand*{\bibnumfmt}[1]{[S#1]}
\renewcommand{\thesubsection}{A\arabic{subsection}}
\setcounter{subsection}{0}

\begin{widetext}

\begin{center}
\textbf{\Large Supplemental Material for}

\vskip6mm

{{\large \bf \noindent
``Topological Correspondence between Hermitian and Non-Hermitian Systems:
Anomalous Dynamics"}}

\vskip3mm

{\noindent
Jong Yeon Lee, Junyeong Ahn, Hengyun Zhou, and Ashvin Vishwanath}

\vskip3mm

\end{center}
\end{widetext}

\tableofcontents

\section{Symmetries and Spectrum}

In this section, we review~\cite{zhou2019periodic,budich2019symmetry} how non-Hermitian symmetries constrain the eigenvalue spectrum. First, consider the $K$-type symmetry defined in Eq.~\eqref{eq:symK}. Let $v_\bk$ be a right eigenvector with eigenvalue $\lambda_\bk$. Then, 
\begin{align}
& h_\bk v_\bk = \lambda_\bk v_\bk = \epsilon_k k h_{-\bk}^* k^\dagger v_\bk \nonumber \\
& \Rightarrow\quad  \epsilon_k \lambda_\bk \qty( k^\dagger v_\bk ) =  h^*_{-\bk} \qty( k^\dagger v_\bk) \nonumber \\
&\Rightarrow\quad \epsilon_k \lambda_{\bk}^*(k^Tv_\bk^*) = h_{-\bk} \qty(k^Tv_\bk^*) 
    %\nonumber \\
    %&\Rightarrow\quad \exists \textrm{ Pair of eigenvalues }(\lambda,\epsilon_k \lambda^*).
\end{align}
Therefore, the $K$-type symmetry implies that there exists a right eigenvector $k^Tv_\bk^*$ for $h_{-\bk}$ with an eigenvalue $\lambda_{-\bk} = \epsilon_k \lambda^*_\bk$. At time reversal invariant momenta (TRIM), it guarantees the existence of a pair of eigenvalues $(\lambda^{ }_\bk, \epsilon_k \lambda_\bk^*)$. 

Second, consider the $C$-type symmetry defined in Eq.~\eqref{eq:symC}. Note that in a non-Hermitian matrix, for a given set of eigenvalues $\{\lambda_n\}$, the set of left eigenvectors $\{u_n\}$ and the set of right eigenvectors $\{v_n\}$ are different in general. Therefore, the {appropriate} generalization of orthonormality is that the left and right eigenvectors form a biorthonormal system, where $u_n v_m = \delta_{n,m}$. Again, let $v_\bk$ be a right eigenvector with eigenvalue $\lambda_\bk$. Then,
\begin{align}
& h_\bk v_\bk = \lambda_\bk v_\bk = \epsilon_c c h^T_{-\bk} c^\dagger v_\bk 
\nonumber \\
&\Rightarrow \quad \epsilon_c \lambda_\bk \qty(c^\dagger v_\bk) = 
h^T_{-\bk} \qty( c^\dagger v_\bk ) \nonumber \\
&\Rightarrow \quad 
(v_\bk^T c^* ) h_{-\bk} = \epsilon_c \lambda_\bk (v_\bk^T c^* )
\end{align}
Therefore, $v^T_\bk c^*$ is now a left eigenvector of $h_{-\bk}$ with an eigenvalue $\lambda_{-\bk} = \epsilon_c \lambda_\bk$. At TRIM, if $\epsilon_c = -1$, it guarantees a pair of eigenvalues $(\lambda_\bk, -\lambda_\bk)$. 

However, when $\epsilon_c = +1$, {the two eigenvalues are the same.} Therefore, if $(v^T_\bk c^*, v_\bk)$ is a biorthonormal pair of left and right eigenvectors, then the TRIM do not have any degeneracies. {Otherwise, there will be a degeneracy, since the biorthonormal partner of  $v^T_\bk c^*$, also a right eigenvector, shares the same eigenvalue as $v_\bk$ but is linearly independent.} This condition can be satisfied when $\eta_c = -1$. To {see this}, note that $c c^* = \eta_c = -1$ implies $c^\dagger = -c^*$. Now, assume that $(v^T_\bk c^*, v_\bk)$ is a biorthonormal pair. Then, we have
\begin{align}
&\lambda = v^T_\bk c^* v_\bk = v_\bk^T c^\dagger v_\bk = v_\bk^T (-c^*) v_\bk = -\lambda,
\end{align}
where we performed a transpose in the second equality and used $c^\dagger = -c^*$ in the third equality. Therefore, $\lambda = 0$, implying that  $v^T_\bk c^*$ and $v_\bk$ are orthogonal to each other. This guarantees the symmetry protected degeneracy, and {constitutes a biorthogonal generalization of the Kramers degeneracy.} Such a degeneracy does not exist when $\epsilon_c = -1$.

Finally, consider a $Q$-type symmetry in Eq.~\eqref{eq:symQ}. Its effect is very similar to a $K$-type symmetry:
\begin{align}
& h_\bk v_\bk = \lambda_\bk v_\bk = \epsilon_q q h_{\bk}^\dagger q^\dagger v_\bk \nonumber \\
& \Rightarrow\quad   \epsilon_q \lambda_\bk \qty( q^\dagger v_\bk ) =  h^\dagger_{\bk} \qty( q^\dagger v_\bk) \nonumber \\
&\Rightarrow\quad \epsilon_q \lambda^*(v_\bk^\dagger q) =  (v_\bk^\dagger q) h_{\bk}
\end{align}
Therefore, the $Q$-symmetry relates the right and left eigenvectors, and  implies that for every $\bk$-point, there is a pair of eigenvalues $(\lambda_\bk, \epsilon_q \lambda_\bk^*)$. In the situation where $\lambda_\bk = \epsilon_q \lambda_\bk^*$, this does not guarantee a degeneracy. For real eigenvalues with $\epsilon_q=-1$, this acts similar to the chiral symmetry of a Hermitian system.

\section{Non-Hermitian Hamiltonian from Lindblad Equation}
In this section, we provide a brief review of how a non-Hermitian Hamiltonian emerges as an effective description of {a subsystem in a full quantum system}. Consider a system coupled to an environment. If the total density matrix is $\rho_\textrm{tot}$, the system density matrix can be obtained by tracing out the environment, $\rho = \Tr_\textrm{E} \rho_\textrm{tot}$. Then, under the Markovian dynamics assumptions, one can derive that the system's density matrix evolves under the following Lindblad (GKSL) master equation:
\begin{equation}\label{eq:master}
    \frac{d \rho}{dt} = -\frac{i}{\hbar}\qty( H_\textrm{eff} \rho -  \rho H_\textrm{eff}^\dagger )  + \sum_m L_m \rho L_m^\dagger
\end{equation}
where $H_\textrm{eff} = H - (i/2) \sum_m L_m^\dagger L_m $,$H$ is the system Hamiltonian and $L_m$ are Lindblad (or jump) operators.
$L_m$ is called a quantum jump because it introduces an abrupt collapse of the state, similar to a measurement. In this form, the second term is often called the recycling term, as it recycles the population that is lost from certain states due to the non-Hermitian effective Hamiltonian, placing it in other states. Indeed, if we inspect the last term, its trace is given by $\sum_m \tr (L_m \rho L_m^\dagger) = \tr (\rho \sum_m L_m^\dagger L_m) > 0$ because $\sum_m L_m^\dagger L_m$ and $\rho$ are positive semidefinite operators. {This acts to increase the trace of the density matrix, which exactly counteracts the loss due to the first term, a non-Hermitian time evolution term under $H_\textrm{eff}$. Here, the imaginary part of the eigenvalues of $H_\textrm{eff}$ are non-positive, as can be seen from the fact that $\sum_m L_m^\dagger L_m$ is always positive semidefinite. Note however that} in the classical context, e.g. non-Hermitian systems arising from photonic crystals with gain and loss, such a constraint does not exist.

As an example, consider a zero-dimensional bosonic system consisting of a single mode of frequency $\omega$. Let $H = \hbar \omega b^\dagger b$, and $L = \gamma b$. Then, $H_\textrm{eff} = \omega b^\dagger b - i \frac{\gamma^2}{2} b^\dagger b$. For simplicity, assume $\omega = 0$ and the initial state was a pure state $\rho_0 = \ket{\psi}\bra{\psi}$ with $\ket{\psi} = 1/\sqrt{2}(\ket{0}+\ket{1})$. If we only consider the first term in Eq.~\eqref{eq:master}, a non-Hermitian time evolution part, we get
\begin{equation}\label{eq:NH_evolution}
    \ket{\psi(t)}  = e^{-i H_\textrm{eff} t} \ket{\psi} \mapsto  \frac{ e^{-\gamma^2t/2}\ket{1} + \ket{0}}{\sqrt{1 + e^{-\gamma^2t}}}
\end{equation}
where the denominator is introduced to normalize the state. {If we can obtain information about the environment, then we can condition our analysis to a case where no quantum jump occurs. For example, we can include only the data where we did not observe an increase of the boson number of the environment.} Then, the time evolution is described by Eq.~\eqref{eq:NH_evolution}. This formalism can be easily generalized to fermionic systems, as in Ref.~\onlinecite{Ashida2018_jump}. 

Another example is the non-Hermitian chiral hopping mode, which can arise as an effective Hamiltonian as well~\cite{gong2018topological}. If we take $H =-t \sum_n \qty( c^\dagger_n c_{n-1} + \textrm{h.c.} )$ and $ L_n=  c_n + i c_{n-1}$, we obtain
\begin{align}
H_\textrm{eff} =& \sum_n \Big( -i c^\dagger_n c_n + (-t - 1/2) c^\dagger_n c_{n-1} \nonumber \\
&+(-t + 1/2) c^\dagger_{n-1} c_{n} \Big).
\end{align}
Inspecting the spectrum, one can show that the imaginary parts of the eigenvalues are always less than or equal to zero.

\section{Non-Hermitian Description of Transfer Matrices}

In this section, we discuss how a non-Hermitian matrix arises as the transfer matrix of the boundary zero mode in a one-dimensional SPT. Consider a $d$-dimensional massive Dirac Hamiltonian given by
\begin{equation}
    H =   \sum_j -i \alpha_j \frac{\rd}{\rd x_j}  + \beta m(x).  
\end{equation}
where $\alpha_j$ and $\beta$ are anticommuting Dirac matrices. Under the presence of other symmetries, e.g. chiral, particle-hole, and time reversal, the possible topology for the mass-term matrix $\beta$ is constrained, and one can classify all topological phases from this. 

Here, we are interested in the boundary modes of the case with $d=1$. Due to the famous Jackiw-Rebbi mechanism \cite{JackiwRebbi}, the boundary between two different topological phases hosts a localized zero mode. To solve this, consider the following Hamiltonian:
\begin{equation}\label{eq:1D_jackiw}
    H = -i \alpha \rd_x  + m \beta, \qquad H \psi = E \psi
\end{equation}
To obtain the zero energy solution, we take $E=0$ such that 
\begin{equation}
-i \rd_x \psi = - m(x) \alpha \beta \psi \,\, \mapsto \,\,  T_x \psi = - m(x) \alpha \beta \psi
\end{equation}
Therefore, the matrix $-m \alpha \beta$ describes how the zero-energy solution changes under the translation $T_x$~\footnote{As a physical Hilbert space consists of normalizable wavefunctions, we should only take decaying solutions.}. In other words, the matrix $\alpha \beta$ is a transfer matrix along the $x$-direction. 

One can immediately notice that $\alpha \beta$ can be non-Hermitian. For example, if $\alpha = \sigma_1$ and $\beta = \sigma_2$, $\alpha \beta = i \sigma_3$ which is non-Hermitian. Knowing that non-Hermitian matrices can appear as a transfer matrix for zero modes, let us analyze possible symmetries of $\alpha \beta$. As the transfer matrix is constructed in terms of the original Hamiltonian, it is natural to examine whether the symmetries of the original Hamiltonian are inherited. Consider a time-reversal symmetry ${\cal T} = U_T {\cal K}$, particle-hole symmetry ${\cal P} = U_P {\cal K}$ and chiral symmetry ${\cal C} = U_C$ where $U$ matrices are unitary transformations. In order to be symmetries of the original Hamiltonian in \eqnref{eq:1D_jackiw}, the matrices $U$s satisfy
\begin{eqnarray}
    U_T \alpha^* U_T^{-1} = - \alpha &\quad&      U_T \beta^* U_T^{-1} = \beta   \nonumber \\
    U_P \alpha^* U_P^{-1} = \alpha &\quad&      U_P \beta^* U_P^{-1} = - \beta  \nonumber \\
    \{U_C, \alpha \} = 0 &\quad&  \{ U_C, \beta \} = 0
\end{eqnarray}
Based on these commutation relations, we are now ready to analyze the symmetries of the ``transfer matrix'' $\alpha \beta$. We can show that $M$ transforms in the following way under Hermitian conjugation, transposition, and complex conjugation with additional unitary transformations $U_{C,T,K}$:
\begin{eqnarray}
    &U_C (\alpha \beta)^\dagger U_C^{-1} = U_C \beta^\dagger \alpha^\dagger U_C^{-1} = U_C \beta \alpha U_C^{-1} = - \alpha \beta \nonumber \\
    &U_T (\alpha \beta)^T U_T^{-1} = U_T \beta^T \alpha^T U_T^{-1} = U_T \beta^* \alpha^* U_T^{-1} = - \beta \alpha = \alpha \beta \nonumber \\
    &U_P (\alpha \beta)^* U_P^{-1} =  U_P \alpha^* \beta^* U_P^{-1} = - \alpha \beta 
\end{eqnarray}
where we used the fact that transposition is equivalent to complex conjugation for $\alpha$ and $\beta$, as they are Hermitian matrices. 

In summary, the symmetries $({\cal C}, {\cal T}, {\cal P})$ of the original 1D Hamiltonian give rise to the following non-Hermitian symmetries $(Q,C,K)$ (pseudo-hermiticity, transpose, and complex conjugation)  respectively for the transfer matrix of the boundary mode. Therefore, if the original Hamiltonian (1D) is in the AZ class $s$, then the non-Hermitian transfer matrix (0D) belongs to the AZ$^\dagger$ class $s^\dagger$. This provides an interesting example where a non-Hermitian matrix naturally arises with symmetries inherited from the Hermitian system in one higher dimension.  Surprisingly, the correspondence exactly agrees with the topological correspondence between Hermitian and non-Hermitian systems proven in the main text. Note however that this example does not generalize easily into higher dimensional cases.

\section{NH classes AII$^\dagger$,  AIII$^\dagger$, and CI$^\dagger$ in 2D}

\begin{figure}
 \hspace{-5pt} \includegraphics[width=0.45\textwidth]{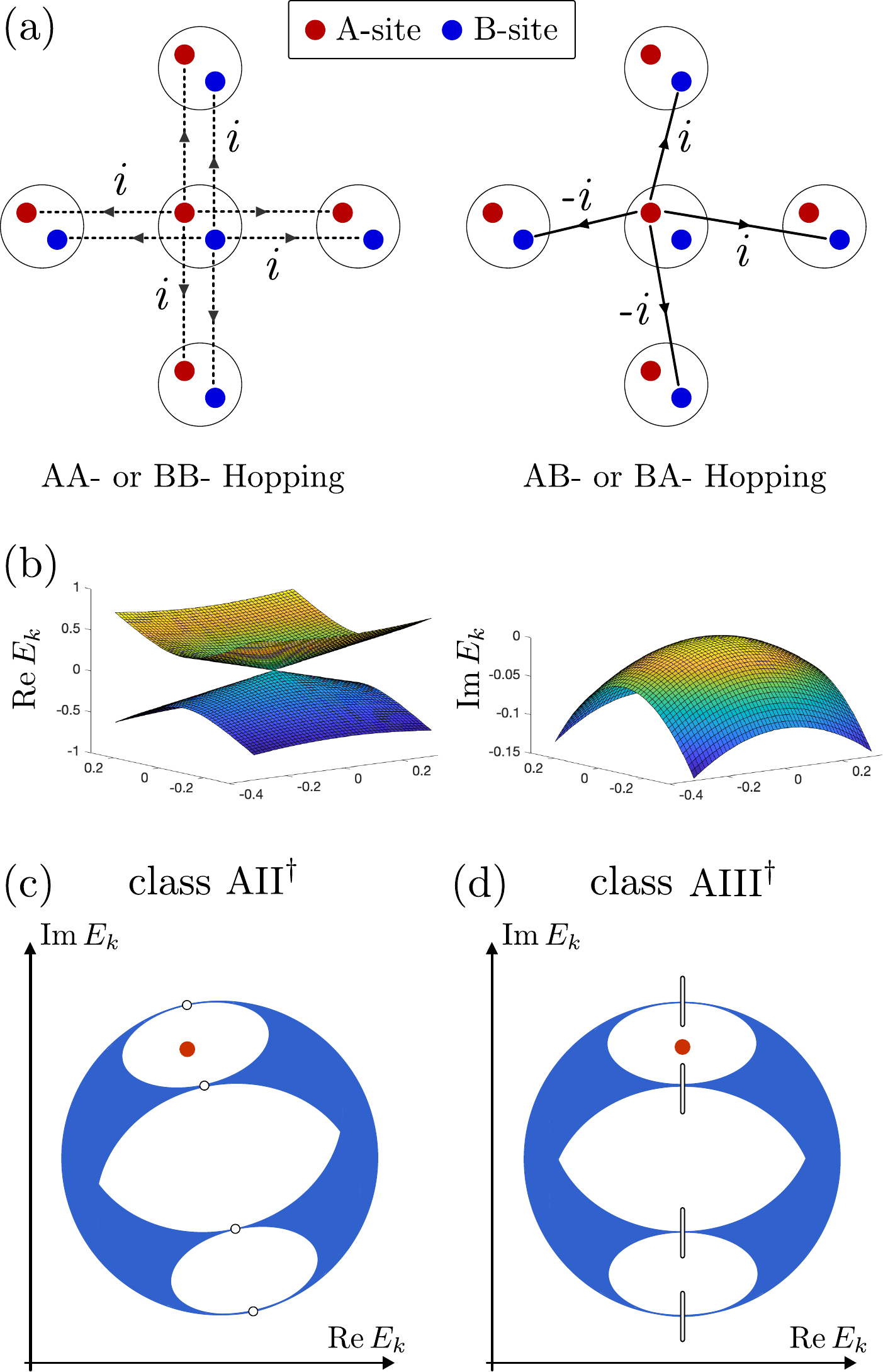}
\caption{ \label{fig:2Dmodel} (a) A non-Hermitian lattice model for the symmetry class NH AII$^\dagger$ with nontrivial topology. Here, $i$ or $-i$ with directions represents the phase structure of the hopping terms; the AB-hopping is non-Hermitian while the AA- or BB- hopings are Hermitian, as one can see from the figure. (b) Real and imaginary part of the dispersion of the model in \eqnref{eq:2D_equation} for generic parameters around $\bk = (0,0)$. (c,d) More generic complex dispersions for class AII$^\dagger$ and AIII$^\dagger$. For AII$^\dagger$, the entire spectrum can be further rotated. For AIII$^\dagger$, Dirac cones can be further deformed to exceptional rings.}
\end{figure}

In this section, we construct a two-dimensional  non-Hermitian system in the class AII$^\dagger$, which corresponds to the boundary of a three-dimensional Hermitian system in class AII (topological insulator). The NH class AII$^\dagger$ has a transpose symmetry with $c = \sigma_2$ so that $c c^* = -1$, which ensures a generalized Kramers degeneracy. The model is given by the following $k$-space Hamiltonian:
\begin{eqnarray}\label{eq:2D_equation}
    H_k &=& i(\gamma_1 \cos k_x + \gamma_2 \cos k_y - \gamma_0) \nonumber \\
    && \quad + t_1 \sin k_x \sigma_1 + t_2 \sin k_y \sigma_2.
\end{eqnarray}
The model Hamiltonian in real space looks rather exotic; the schematic hopping structure is illustrated in Fig.~\ref{fig:2Dmodel}. The real space Hamiltonian is given by
\begin{eqnarray}
    H &=& \sum_{r, \eta = A,B} \Big( i \frac{\gamma_1}{2} c_{\eta,r}^\dagger c^{ }_{\eta,r-\hat{x}} + i \frac{\gamma_1}{2} c_{\eta,r-\hat{x}}^\dagger c^{ }_{\eta,r}  \nonumber \\
    && \qquad + i \frac{\gamma_2}{2} c_{\eta,r}^\dagger c^{ }_{\eta,r-\hat{y}} + i \frac{\gamma_2}{2} c_{\eta,r-\hat{y}}^\dagger c^{}_{\eta,r} - i\gamma_0 c^\dagger_{\eta,r} c^{}_{\eta,r}  \Big)  \nonumber \\
    && + \sum_r \Big( i t_1 c_{A,r}^\dagger c^{ }_{B,r-\hat{x}} + i t_1 c_{B,r}^\dagger c^{ }_{A,r-\hat{x}}  \nonumber \\
    && \qquad + i t_2 c_{A,r}^\dagger c^{ }_{B,r-\hat{y}} + i t_2 c_{B,r}^\dagger c^{ }_{A,r-\hat{y}}  + \textrm{h.c.} \Big) \qquad 
\end{eqnarray}
The complex dispersion of the system is very similar to Fig.~1(b) of the main text. {Note that this particular model has a pseudo-hermiticity as well, which can be removed by adding additional terms with the correct symmetries. If the pseudo-hermiticity is preserved, then the model has also the complex-conjugation symmetry with $\eta_k = -1$; thus it will belong to class CII$^\dagger$.}

Consider the following generic Hamiltonian in $\bk$-space:
\begin{equation}
    h(\bk) = i \gamma(\bk) + f_1 (\bk) \sigma_1 + f_2(\bk) \sigma_2 +  f_3 (\bk) \sigma_3.
\end{equation}
If $f_i(\bk)$ is an (complex) odd function of $\bk = (k_x,k_y)$ and $\gamma(\bk)$ is an (complex) even function of $\bk$, the Hamiltonian belongs to class AII$^\dagger$. Moreover, as the transpose symmetry does not constrain the phase factor, $\gamma(\bk)$ and $f_i(\bk)$ can generically be complex-valued. In other words, the entire spectrum can be rotated by an arbitrary angle in the complex plane, simply by multiplying $e^{i\theta}$ to \eqnref{eq:2D_equation}, as in \figref{fig:2Dmodel}(c). This is very crucial: it means that the spectrum can be inverted, implying that the net chirality of the Dirac cones can be flipped to give a different invariant. This agrees with the physics of the boundary of a 3D topological insulator, at which the chirality of the Dirac cone is not well-defined, unlike chiral topological order.

However, for the class AIII$^\dagger$, $f_{1,2}(\bk)$ should be real and $f_3(\bk)$ should be imaginary, due to the pseudo-hermiticity ($q=\sigma_3$). With this constraint, they can be any function of $\bk$. {With the $f_3(\bk)$ term, the Dirac cone can be deformed into an exceptional ring, as shown in \figref{fig:2Dmodel}(d).}

To further demonstrate an explicit model and correspondence, consider a non-Hermitian class CI$^\dagger$, whose two-dimensional classification is given by $2\mathbb{Z}$. In the table of the main text (or \tabref{ref:ptab1}), we denoted it by $\mathbb{Z}$, which is isomorphic to $2\mathbb{Z}$; however, $2\mathbb{Z}$ is more precise since the system is characterized by an even number of symmetry-protected gapless boundary modes. The system in class CI$^\dagger$ has $K$ and $C$ type symmetries with $k k^* = -1$ and $c c^* = 1$. Due to the $K$-type symmetry with $k k^* = -1$, one can show that the Dirac cone must be doubly degenerate. This can be achieved by the following two-dimensional quadratic Hamiltonian:  $H_\textrm{herm}=(k_x^2-k_y^2)\sigma_1+2k_xk_y\sigma_3$ with $k=\sigma_2$ and $c=1$. Although the system realizes a single gapless excitation, as the dispersion is quadratic, the net winding number is equivalent to that of two  Dirac cones. A corresponding lattice-regularized non-Hermitian Hamiltonian would be given by, for example,
\begin{eqnarray}
    H &=& i (\cos k_x + \cos k_y) + 2(\cos k_y - \cos k_x) \sigma_1 \nonumber \\
    &&+ (\cos(k_x - k_y) - \cos(k_x + k_y) ) \sigma_2.
\end{eqnarray}
For this Hamiltonian, quadratic gapless points exist at $\bk = (0,0)$ and $(\pi,\pi)$, but only the one at $(0,0)$ survives at long times.
There is another way to construct a Hamiltonian with a linear dispersion, by using a four-dimensional Hamiltonian. Then, the Hermitian part can be written as $H_\textrm{herm} = k_x \sigma_0 \otimes \tau_1 + k_y \sigma_0 \otimes \tau_3$ with $k = \sigma_{2}$ and $c = \sigma_2 \tau_{2}$. 
In either case, one can show that the gapless modes with net winding number two cannot be gapped out due to the symmetry (they can however be deformed into exceptional ring, as pointed out in the main text). This agrees with the boundary of a three-dimensional topological superconductor in Hermitian class CI, which is characterized by an even number of Dirac cones.

\begin{figure} \includegraphics[width=0.47\textwidth]{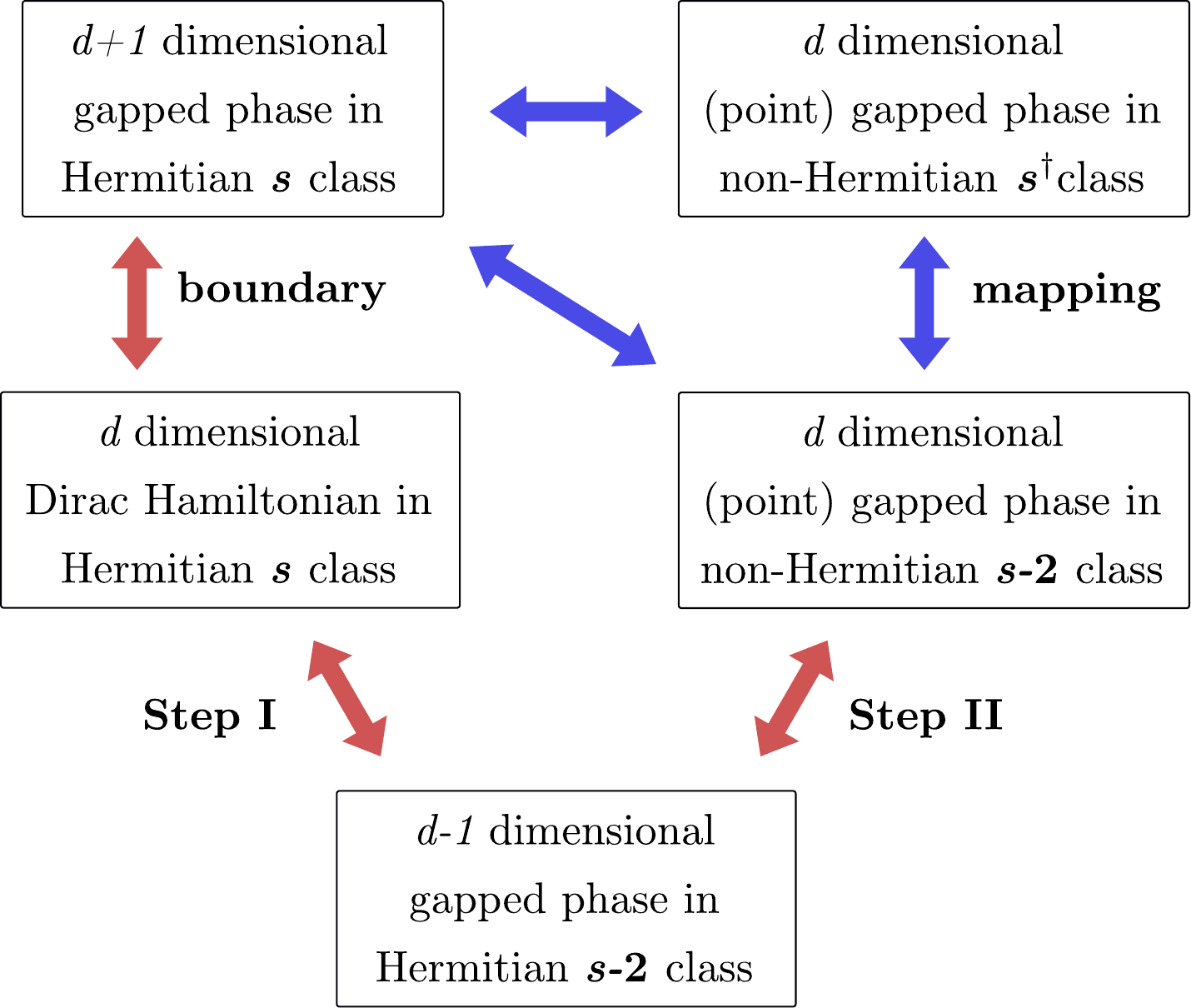}
\caption{ \label{fig:proof_diagram} This diagram illustrates how we prove the correspondence between $(d+1)$-dimensional Hermitian system in AZ class $s$ and $d$-dimensional non-Hermitian system in NH AZ class $s-2$. It also provides an indirect way to prove correspondence for NH AZ$^\dagger$ classes as well.  }
\end{figure}

\section{Proof Part II For Non-Hermitian AZ classes}

In this section, we will prove the  correspondence between anomalous boundary states of the $(d+1)$-dimensional Hermitian class $s$ and bulk states of the $d$-dimensional non-Hermitian class $s-2$, as we have done in the main text for non-Hermitian class $s^{\dagger}$. The procedure is outlined in \figref{fig:proof_diagram}.

{\bf Step I:} Let us first recall that the anomalous boundary state of the $(d+1)$-dimensional Hermitian system is described by the $d$-dimensional Dirac Hamiltonian:
\begin{align}
H_{\rm Dirac}({\bf k})
&={\bf \delta k}\cdot \boldsymbol{\Gamma}
=|{\bf \delta k}| H_{d-1}(\theta_1,...,\theta_{d-1}),
\end{align}
where $\boldsymbol{\Gamma} = (\Gamma_1, ..., \Gamma_d)$ are $d$ anticommuting gamma matrices and we introduced spherical coordinates $-\pi\le \theta_1<\pi$ and $-\pi/2\le \theta_{i=2,...,d-1}\le \pi/2$:
\begin{align}
\delta k_{d}
&=|{\bf \delta k}|\sin\theta_{d-1},\notag\\
\delta k_{d-1}
&=|{\bf \delta k}|\cos\theta_{d-1}\sin\theta_{d-2},\notag\\
&\vdots \notag\\
\delta k_{n}
&=|{\bf \delta k}|\cos\theta_{d-1}\cos\theta_{d-2}\dots \sin\theta_{n-1},\notag\\
&\vdots \notag\\
\delta k_2
&=|{\bf \delta k}|\cos\theta_{d-1}\cos\theta_{d-2}\cos\theta_{d-3}\dots \sin\theta_1,\notag\\
\delta k_1
&=|{\bf \delta k}|\cos\theta_{d-1}\cos\theta_{d-2}\cos\theta_{d-3}\dots \cos\theta_1.
\end{align}
We impose  $T$ (time-reversal) and $P$ (particle-hole) symmetries on the Dirac Hamiltonian.
\begin{align}
TH^*_{{\rm Dirac}}({\bf k})T^{-1}
&=H_{{\rm Dirac}}(-{\bf k}),\notag\\
P H^*_{{\rm Dirac}}({\bf k}) P^{-1}
&=-H_{{\rm Dirac}}(-{\bf k}),
\end{align}
Then, we can prove that $H_{d-1}$ has the following symmetries:
\begin{align}
&(\Gamma_1T) \abs{\delta \bf k} H^*_{d-1}(\vec{ \theta})(\Gamma_1T)^{-1} = (\Gamma_1T)H^*_\textrm{Dirac}({{\bf k}})(\Gamma_1T)^{-1} \nonumber \\
&=- \delta k_1 \Gamma_1 + \sum_{i=2}^{d} \delta k_i \Gamma_i = -\abs{\delta k}  H_{d-1}(-{\vec{\theta}}),\nonumber \\
&\,\,\, \Rightarrow \,  (\Gamma_1T) H^*_{d-1}({\vec{\theta}})(\Gamma_1T)^{-1}  = -H_{d-1}(-{\vec{\theta}}). 
\end{align}
Similarly, one can show that
\begin{equation}
     (\Gamma_1 P)H^*_{d-1}({\vec{\theta}})(\Gamma_1 P)^{-1} = H_{d-1}(-{\vec{\theta}}),
\end{equation}
where ${\vec{\theta}}=(\theta_1,...,\theta_{d-1})$, and
\begin{align}
(\Gamma_1T)(\Gamma_1T)^*
&=-TT^*,\notag\\
(\Gamma_1 P)(\Gamma_1 P)^*
&=+CC^*.
\end{align}
Now $\Gamma_1 T$ has become the particle-hole symmetry operator, and $\Gamma_1 P$ has become the time-reversal symmetry operator. Accordingly, the topological information (anomalousness) of $H_{\rm Dirac}$ in the class $s$ is encoded in a gapped Hamiltonian $H_{d-1}$ in the class $s-2$ and vice versa.

{\bf Step II: } We have shown that the class $s$ $d$-dimensional Hermitian Dirac Hamiltonian maps to a class $s-2$ $(d-1)$-dimensional gapped Hamiltonian. Now, we want to reach the class $s-2$ $(d-1)$-dimensional gapped Hamiltonian from the other side, a $d$-dimensional system in the non-Hermitian class $s-2$ and establish the correspondence between a Hermitian class $s$ and non-Hermitian class $s-2$. Note that the correspondence for class $s^{\dagger}$ in the main text follows from the aforementioned correspondence by the mapping discussed in \ref{sec.duality}. First, we construct the doubled Hamiltonian $\bar{\cal H}$ for ${\cal H}$:
\begin{align}\label{eq:double_sym_def}
\bar{\cal H}
=
\begin{pmatrix}
0&{\cal H}\\
{\cal H}^{\dagger}&0
\end{pmatrix}.
\end{align}
It satisfies (see main text for the definition of the non-Hermitian AZ class)
\begin{align} \label{eq:doubled_condition}
\bar{k}\bar{H}^*({\bf k})\bar{k}^{-1}
&=\bar{H}(-{\bf k}),\notag\\
\bar{c}\bar{H}^*({\bf k})\bar{c}^{-1}
&=-\bar{H}(-{\bf k}),\notag\\
\Sigma\bar{H}({\bf k})\Sigma^{-1}
&=-\bar{H}({\bf k}),
\end{align}
where
\begin{align} 
\bar{k}
=
\begin{pmatrix}
k&0\\
0&k
\end{pmatrix},\quad
\bar{c}
=
\begin{pmatrix}
0&{c}\\
{c}&0
\end{pmatrix},\quad
\Sigma
=
\begin{pmatrix}
1&0\\
0&-1
\end{pmatrix},\notag\\
\bar{k}\bar{k}^*
={k}{k}^*
\begin{pmatrix}
1&0\\
0&1
\end{pmatrix},\quad
\bar{c}\bar{c}^*={c}{c}^*
\begin{pmatrix}
1&0\\
0&1
\end{pmatrix}.
\end{align}

We wish to perform a dimensional reduction, meaning that we want to find a $(d-1)$-dimensional system that encodes all topological information of the current system. We will follow the procedure in Ref.~\cite{teo2010topological} to prove the desired result. Let us consider the $d$-dimensional Brillouin zone as a $d$-sphere parametrized by momentum-like spherical coordinates $-\pi\le k_1<\pi$ and $-\pi/2\le k_{2,...,d}\le \pi/2$, i.e., ones that flip sign under $\bar{k}$ and $\bar{c}$.
Now, we want to remove the dimension along $k_d$.
Let us introduce $\theta\equiv k_d$ to emphasize that this is the parameter to be reduced.
The $d$-th homotopy class of $\bar{\cal H}$ is determined by two sectors: one is the $(d-1)$th homotopy class of it at the $\theta=0$ boundary, and the other is the homotopy class of the Hamiltonian in the region $0<|\theta|<\pi/2$ with $\theta=0$ identified to a point.
Since we want to encode the full information of the $d$-dimensional topology in the $(d-1)$-dimensional manifold, we remove the contribution from the second sector by minimizing the dispersion along $\theta$.
To do so, we introduce an artificial action
\begin{align}
S=\int d{\bf k}d\theta {\rm Tr}[\partial_{\theta}\bar{\cal H}\partial_{\theta}\bar{\cal H}].
\end{align}
If we spectrally flatten $\bar{\cal H}$ so that $\bar{\cal H}^2=1$, the Euler-Lagrange equation becomes
\begin{align}
\partial_{\theta}^2\bar{\cal H}+\bar{\cal H}=0,
\end{align}
and the solution is
\begin{align}
\bar{\cal H}({\bf k},\theta)=\cos\theta \bar{H}_{d-1}({\bf k})+\sin\theta \bar{V},
\end{align}
where ${\bf k}=(k_1,...,k_{d-1})$, and $\bar{V}$ is constant. The above Hamiltonian should satisfy the conditions in Eq.~\eqref{eq:doubled_condition}. 
$\bar{\cal H}^2=1$ requires that $\bar{H}_{d-1}^2=\bar{V}^2=1$ and $\{\bar{H}_{d-1},\bar{V}\}=0$.
Due to the additional chiral symmetry $\Sigma$ (on top of $\bar{\Gamma} \equiv \bar{k} \bar{c}$),  $\Sigma\bar{\cal H}\Sigma^{-1}=-\bar{\cal H}$, where $\Sigma=\tau_z$, both $\bar{H}_{d-1}$ and $\bar{V}$ are off-diagonal in the $\tau_z$ eigenbasis.
Let us take $\bar{V}=-\tau_y$, then $\bar{H}_{d-1}$ takes the form of $H_{d-1}\tau_x$.
Thus, we have continuously deformed ${\cal H}$ into the form
\begin{align}
{\cal H}(k_1,...,k_{d-1},\theta)
&=\cos\theta H_{d-1}({\bf k})+i\sin\theta I,
\end{align}
where $H_{d-1}$ is a gapped Hermitian Hamiltonian in $(d-1)$ dimensions, and $I$ is the identity matrix.

As $\cal H$ is in the class $s-2$, $H_{d-1}({\bf k})$ is also in the class $s-2$.
The Hermitian part of $\cal H$ vanishes at $\theta=\pm \pi/2$, so, up to linear order in $\delta \theta$,
\begin{align}
{\cal H}(k_1,...,k_{d-1},\pm \pi/2+\delta \theta)
=\mp \delta \theta H_{d-1}({\bf k})\pm i I.
\end{align}
Notice that $\delta \theta H_{d-1}$ describes a gap-closing point at $\delta\theta=0$ whose winding number $n$ is given by the $(d-1)$th homotopy class of $H_{d-1}$.
When $H_{d-1}$ is spectrally flattened, the gap-closing point becomes a $n$-fold Dirac point.
As $H_{d-1}$ is in the class $s-2$, the Dirac Hamiltonian is in the class $s$ as we have shown above.
This shows that each of the two gap-closing points of the Hermitian part at $\theta=\pm \pi/2$ corresponds to the anomalous boundary states of a $(d+1)$-dimensional topological insulator in the class $s$. Therefore, we can conclude that only the states with $\theta=\theta_0$ ($\theta=-\theta_0$) will survive because of the damping/gain given by the non-Hermitian term $\pm i I$ at long times. 

Note that this does not mean that such an anomalous theory interpretation holds for {any realizations of the model.} Rather, as we saw in the intermediate step, {the statement is that} if the non-Hermitian topology is nontrivial, we can always adiabatically deform the corresponding system to realize anomalous dynamics at long times.

\section{Proof Part III}

In the main text, we have shown by explicit construction that a non-Hermitian system with nontrivial topology can exhibit emergent anomalous dynamics. Conversely, using the Brouwer's fixed-point theorem, one can prove that if a non-Hermitian system exhibits emergent anomalous dynamics, it necessarily implies a nontrivial non-Hermitian point-gap topology.

Let us consider a Dirac point appearing in the long time scale, i.e., the effective Hamiltonian is given by $h=\sum_{n=1}^d k_n\Gamma_n + i \gamma_0$ with $\gamma_0 > 0$, where the last term ensures that this Dirac Hamiltonian dynamics survives at long times.
We can assume without loss of generality that only one Dirac point appears.
In this case, we get a nontrivial topological phase if we choose the complex base point $E_B$ on the imaginary axis such that $\gamma_0 > \Im E_B$ and $\Im E_B$ is larger than the imaginary energy of any other state on the imaginary axis.

With this choice of the base point, let us deform the doubled Hamiltonian into the massive Dirac form: $\bar{H}=\sum_{\mu=0}^{d}f_{\mu}\Gamma_{\mu}$ (that is, we deform to have only mutually anticommuting gamma matrix terms).
Also, we deform it further so that its eigenvalues are $\pm 1$.
Then, we have a flattened Hamiltonian $\bar{Q}=(1/\abs{\bf f})\sum_{\mu=0}^{d}f_{\mu}\Gamma_{\mu}$.
Notice that $\bar{Q}({\bf k}):(k_1,...,k_d)\rightarrow(f_0,..,f_d)/\abs{\bf f}$ is a map from the $d$-dimensional Brillouin torus to a $d$-dimensional sphere $\bar{Q}:T^d\rightarrow S^d$.

Suppose that the Jacobian determinant of the map, $\det D\bar{Q}$, where $\qty( D \bar{Q} )_{ij} = \rd \bar{Q}_i/ \rd k_j$, is nonvanishing everywhere in the Brillouin zone.
Then, the winding number(=degree of the map=$\deg\bar{Q}$) of such a map can be simply evaluated using the Brouwer fixed-point theorem. According to the theorem, the winding number can be obtained by counting the number of ${\bf k}$ points at which the flattened doubled Hamiltonian $\bar{Q}$ is mapped to a Hamiltonian $y$.
More concretely, $\deg_{y}(\bar{Q})=\sum_{x\in \bar{Q}^{-1}(y)}{\rm sgn}\qty(\det D\bar{Q}(x))$.
Here, $y$ can be chosen arbitrarily because the winding number does not depend on this choice.
We take $y=i\Gamma_0$.
Recall that there is only one momentum that is mapped to $i\Gamma_0$ by construction, and the momentum is the location of the Dirac point that survives in the long time evolution.
Accordingly, we have $\deg(\bar{Q})=\pm 1$.

Generically, we must also consider the constraints from time reversal or particle-hole symmetry on the map $\bar{Q}$.
{This simply corresponds to considering the homotopy class on half the Brillouin zone, and will not change the conclusion.}

\section{Mapping between AZ and AZ$^{\dagger}$ classes}
\label{sec.duality}

In this section, we prove the correspondence between the NH AZ$^{\dagger}$ and NH AZ classes.
We will show that we can transform a class $s-2$ non-Hermitian Hamiltonian to a class $s^{\dagger}$ Hamiltonian by multiplying a constant matrix, and vice versa. For the NH AZ class, particle hole symmetry is given by the $C$-type transposition symmetry $c H^T_{\bf k}c^{-1}=\epsilon_c H_{-{\bf k}}$ with $\epsilon_c = -1$. Then,
\begin{align}
c^TH^T({\bf k})=\epsilon_c\eta_c H(-{\bf k})c,
\end{align}
because $c^T=\eta_c c$.
Let us define $\tilde{H}({\bf k})= H({\bf k})c M^\dagger$, where 
$M$ is an unitary operator chosen in such a way that $M M^*
=-c c^*=-\eta_c$. This is always possible in the classification problem, where trivial bands (enlarging the Hilbert space) can be added freely. For example, we can take $M=\sigma_y$ when $\eta_c=1$ and $M=1$ when $\eta_c=-1$
We then have
\begin{align}
M\tilde{H}^T({\bf k})M^{-1}
&=-\epsilon_c\tilde{H}(-{\bf k}), \quad M M^* = -\eta_c
\end{align}
Therefore, for example, if $H$ belonged to classes NH C ($s=6$) or NH D ($s=2$), the new Hamiltonian $\tilde{H}$ belongs to NH AI$^\dagger$ ($s=0$) or NH AII$^\dagger$ ($s=4$).

Next, for NH AZ classes, time reversal symmetry is given by the $K$-type complex-conjugation symmetry, $k H_{\bf k}^* k^{-1} = \epsilon_k H_{-\bf k}$ with $\epsilon_k = +1$. If this is the only symmetry, it is very easy to change the sign for $\epsilon$ by taking $\tilde{H} = i H$. Therefore, for example, if $H$ belonged to  classes NH AI ($s=0$) or NH AII ($s=4$), the new Hamiltonian $\tilde{H}$ belongs to NH D$^\dagger$ ($s=2$) or NH C$^\dagger$ ($s=6$).

Finally, consider the case where we have both time reversal and particle hole symmetries. Define $\tilde{H}({\bf k})= e^{i\phi} H({\bf k})c M^\dagger$. First, this newly defined Hamiltonian satisfies the transformation $(\epsilon_c, \eta_c) \mapsto (-\epsilon_c, -\eta_c)$. To prove the mapping, we also need to show that $(\epsilon_k,\eta_k) \mapsto (-\epsilon_k, \eta_k)$. It holds that
\begin{align}
k\tilde{H}^*({\bf k})k^{-1}
&=e^{-i\phi} k H^*({\bf k}) c^* M^{T} k^{-1}\notag\\
&=e^{-i\phi} k H^*({\bf k}) k^{-1} \qty( k c^*  k^{-1} ) \qty( k M^T k^{-1} ) \notag\\
&=e^{-i\phi} \epsilon_k \epsilon_{kc} \epsilon_{kM} H(-{\bf k}) c M^\dagger \notag \\
&= e^{-2i\phi}\epsilon_k \epsilon_{kc} \epsilon_{kM} \tilde{H} (-{\bf k}),
\end{align}
where $\epsilon_{kc}$ and $\epsilon_{kM}$ are defined by
\begin{align}
\label{eq.rel}
k c^* k^{-1} = \epsilon_{kc} c, \qquad  k M^* k^{-1} = \epsilon_{kM} M.
\end{align}
The first relation {is specified from the commutation relation} between $K$ and $C$ type symmetries \cite{zhou2019periodic}. The second relation {specifies our choice of }matrix $M$, such that $(i)$ $M M^* = - \eta_c$ and $(ii)$ $k M^* k^{-1} = \epsilon_{kM} M$ for some $\epsilon_{kM}$ of unit modulus. Then, by choosing a proper $\phi$, we can always map $\epsilon_k \mapsto -\epsilon_k$. This completes the mapping.

At the level of the doubled Hamiltonian, the classification equivalence between NH classes ($s-2$) and s$^\dagger$ can be easily derived. For the doubled Hamiltonian $\bar{H}$,
\begin{align}
\bar{c}
\bar{H}^*({\bf k})
\bar{c}^{-1}
=\epsilon_C\bar{H}({-\bf k}),\quad
\bar{c}\bar{c}^*=\eta_c.
\end{align}
with
\begin{align}
\bar{H}=
\begin{pmatrix}
0&H\\
H^{\dagger}&0
\end{pmatrix}, \quad 
\bar{c}
=
\begin{pmatrix}
0&c\\
c&0
\end{pmatrix}, \quad 
\Sigma=
\begin{pmatrix}
1&0\\
0&-1
\end{pmatrix},
\end{align}
where $\Sigma$ is an emergent chiral symmetry satisfying \begin{align}
\Sigma
\bar{H}({\bf k})
\Sigma^{-1}
=-\bar{H}({\bf k}),
\end{align}
{Then, the doubled Hamiltonian has a $C$-type symmetry with unitary implementation $\bar{c}\Sigma$, with an opposite signs of $\epsilon_c$:}
\begin{align}
(\bar{c}\Sigma)
\bar{H}^*({\bf k})
(\bar{c}\Sigma)^{-1}
=-\epsilon_c\bar{H}(-{\bf k}),\quad
(\bar{c}\Sigma)(\bar{c}\Sigma)^*=-\eta_c.
\end{align}
Since a doubled Hamiltonian with $\bar{c}$ and $\Sigma$ symmetries can be regarded as a doubled Hamiltonian with $\bar{c}\Sigma$ and $\Sigma$ symmetries, there is a $C$-type mapping $\bar{c}\leftrightarrow \bar{c}\Sigma$ with $\epsilon_c,\eta_c\leftrightarrow -\epsilon_c,-\eta_c$ for doubled Hamiltonians. For $K$-type mapping, one can show that $\bar{k} \Sigma$ satisfies (see \eqnref{eq:double_sym_def} for the definition of $\bar{k}$)
\begin{align}
(\bar{k}\Sigma)
\bar{H}^*({\bf k})
(\bar{k}\Sigma)^{-1}
=-\epsilon_k\bar{H}(-{\bf k}),\quad
(\bar{k}\Sigma)(\bar{k}\Sigma)^*=\eta_k.
\end{align}
where the difference between the type-$C$ symmetry comes from the fact that $\bar{k}$ and $\Sigma$ commute rather than anticommute. Therefore, at the doubled Hamiltonian level, by mapping $\bar{c} \mapsto \bar{c} \Sigma$ and $\bar{k} \mapsto \bar{k} \Sigma$, $(\eta_c,\epsilon_c,\eta_k,\epsilon_k) \mapsto (-\eta_c, -\epsilon_c, \eta_k, -\epsilon_k)$, meaning that NH classes $s^\dagger$ and $s-2$ have the same point-gap classification.

\section{Non-Hermitian Bernard-LeClair 38 symmetry classes}

In the classification framework, we map a non-Hermitian system into a Hermitian one by doubling. Meanwhile, both $C$ and $K$ symmetries become antiunitary symmetries at the doubled level, which may seem to obscure the distinction between the NH AZ and NH AZ$^\dagger$ classes. This is related to the redundancy associated with the redefinition of the given symmetries~\cite{zhou2019periodic,kawabata2018symmetry}. However,  the way doubled symmetries are implemented distinguish two classes. At the doubled level, $\bar{c} = (\sigma_x \otimes c) {\cal K}$ and $\bar{k} = (\mathbb{I}_2 \otimes k)  {\cal K}$, which gives different commutation relations with an additional chiral symmetry of the doubled Hamiltonian, $\Sigma = \sigma_z \otimes \mathbb{I}$. 

We now discuss how previously-studied non-Hermitian 38-fold Bernard-LeClair symmetry classes \cite{zhou2019periodic,kawabata2018symmetry} are mapped into NH AZ and AZ$^\dagger$ classes. Because complex conjugation symmetry is defined by $k = qc$ (or $cq^*$), we can show that $k k^* = \eta_k = \eta_c \epsilon_{qc}$. In Ref.~\cite{zhou2019periodic}, we always set $q h^\dagger q^{-1} = h$. Therefore, in this NH AZ$^\dagger$ case where $k h^* k = -h$, and $q h^\dagger q^{-1} = - h$, we have to map $h \mapsto i h$ such that $(-)$ signs are all removed. This would rotate the spectrum, but would not change its \emph{shape}. However, since real and imaginary parts of the energy  have totally different physical meanings, two different representations with $\epsilon_q = +1$ and $-1$ should be considered differently, as emphasized in TABLE.~\ref{ref:ptab1}
%It is an interesting question whether this is allowed if multiplying $h$ by unitary matrix is not allowed. In terms of level statistics, this might be completely meaningless. 
In TABLE.~\ref{ref:ptab1} and \ref{ref:ptab2}, we summarize how NH AZ and AZ$^\dagger$ classes are mapped to symmetry classes in Ref.~\cite{zhou2019periodic}.

\newpage

\begin{table*}\begin{center}\begin{tabular}{ | c | c | c | c | c | c | c | c | c |}\hline Sym. & Gen. Rel. & Clifford Generators & Cl. Sp. 0D & $d = 0$ & 1 & 2 & 3 & Known Class\\\hline1,  &  & $\{\gamma ,m,\Sigma \}$ &  $\mathcal{C}_1$ & $0$ & $\mathbb{Z}$ & $0$ & $\mathbb{Z}$ & NH A$^{(\dagger)}$\\\hline2, P &  & $\{\gamma ,m,\Sigma \}_{\Sigma P}$ &  $\mathcal{C}_1^{\times 2}$ & $0$ & $\mathbb{Z}^{\times 2}$ & $0$ & $\mathbb{Z}^{\times 2}$ &  \\
\hline3, Q & \begin{tabular}{c}$\epsilon_q=1$\\\hline$\epsilon_q=-1$\end{tabular} & $\{\gamma ,m\}_{Q}$ &  $\mathcal{C}_1$ & $0$ & $\mathbb{Z}$ & $0$ & $\mathbb{Z}$ & \begin{tabular}{c} A \\ \hline NH AIII$^{(\dagger)}$ \end{tabular} \\
\hline4, PQ & $\epsilon_{pq}=1$ & $\{\gamma ,m,P\}_{Q}$ &  $\mathcal{C}_1$ & $0$ & $\mathbb{Z}$ & $0$ & $\mathbb{Z}$ & AIII\\\hline5, PQ & $\epsilon_{pq}=-1$ & $\{\gamma ,m\}_{Q,\Sigma P}$ &  $\mathcal{C}_0^{\times 2}$ & $\mathbb{Z}^{\times 2}$ & $0$ & $\mathbb{Z}^{\times 2}$ & $0$ & 
\\\hline6, C & $\epsilon_c=1$, $\eta_c=1$ & $\{\gamma ,Jm,\Sigma ,C,JC\}$ &  $\mathcal{R}_7$ & $0$ & $0$ & $0$ & $\mathbb{Z}$ & NH AI$^\dagger$
\\\hline7, C & $\epsilon_c=1$, $\eta_c=-1$ & $\{\gamma ,Jm,\Sigma ,C,JC\}$ &  $\mathcal{R}_3$ & $0$ & $\mathbb{Z}_2$ & $\mathbb{Z}_2$ & $\mathbb{Z}$ & NH AII$^\dagger$
\\\hline8, C & $\epsilon_c=-1$, $\eta_c=1$ & $\{J\gamma ,m,\Sigma ,C,JC\}$ &  $\mathcal{R}_3$ & $0$ & $\mathbb{Z}_2$ & $\mathbb{Z}_2$ & $\mathbb{Z}$ & NH D
\\\hline9, C & $\epsilon_c=-1$, $\eta_c=-1$ & $\{J\gamma ,m,\Sigma ,C,JC\}$ &  $\mathcal{R}_7$ & $0$ & $0$ & $0$ & $\mathbb{Z}$ & NH C
\\\hline10, PC & \begin{tabular}{c}$\epsilon_c=1$, $\eta_c=1$, $\epsilon_{pc}=1$\\\hline$\epsilon_c=-1$, $\eta_c=1$, $\epsilon_{pc}=1$\end{tabular} & \begin{tabular}{c}$\{\gamma ,Jm,\Sigma ,C,JC\}, [J\Sigma P]$\\\hline$\{J\gamma ,m,\Sigma ,C,JC\}, [J\Sigma P]$\end{tabular} &  $\mathcal{C}_1$ & $0$ & $\mathbb{Z}$ & $0$ & $\mathbb{Z}$ & \\ \hline11, PC & \begin{tabular}{c}$\epsilon_c=1$, $\eta_c=-1$, $\epsilon_{pc}=1$\\\hline$\epsilon_c=-1$, $\eta_c=-1$, $\epsilon_{pc}=1$\end{tabular} & \begin{tabular}{c}$\{\gamma ,Jm,\Sigma ,C,JC\}, [J\Sigma P]$\\\hline$\{J\gamma ,m,\Sigma ,C,JC\}, [J\Sigma P]$\end{tabular} &  $\mathcal{C}_1$ & $0$ & $\mathbb{Z}$ & $0$ & $\mathbb{Z}$ & \\\hline12, PC & \begin{tabular}{c}$\epsilon_c=1$, $\eta_c=1$, $\epsilon_{pc}=-1$\\\hline$\epsilon_c=-1$, $\eta_c=-1$, $\epsilon_{pc}=-1$\end{tabular} & \begin{tabular}{c}$\{\gamma ,Jm,\Sigma ,C,JC\}_{\Sigma P}$\\\hline$\{J\gamma ,m,\Sigma ,C,JC\}_{\Sigma P}$\end{tabular} &  $\mathcal{R}_7^{\times 2}$ & $0$ & $0$ & $0$ & $\mathbb{Z}^{\times 2}$ & \\\hline13, PC & \begin{tabular}{c}$\epsilon_c=1$, $\eta_c=-1$, $\epsilon_{pc}=-1$\\\hline$\epsilon_c=-1$, $\eta_c=1$, $\epsilon_{pc}=-1$\end{tabular} & \begin{tabular}{c}$\{\gamma ,Jm,\Sigma ,C,JC\}_{\Sigma P}$\\\hline$\{J\gamma ,m,\Sigma ,C,JC\}_{\Sigma P}$\end{tabular} &  $\mathcal{R}_3^{\times 2}$ & $0$ & $\mathbb{Z}_2^{\times 2}$ & $\mathbb{Z}_2^{\times 2}$ & $\mathbb{Z}^{\times 2}$ & \\\hline14, QC & \begin{tabular}{c} $\epsilon_q = 1$, $\epsilon_c=1$, $\eta_c=1$, $\epsilon_{qc}=1$ \\ \hline $\epsilon_q = -1$, $\epsilon_c=1$, $\eta_c=1$, $\epsilon_{qc}=1$ \end{tabular} & $\{\gamma ,Jm,C,JC\}_{Q}$ &  $\mathcal{R}_0$ & $\mathbb{Z}$ & $0$ & $0$ & $0$ & \begin{tabular}{c}AI  \\ \hline NH BDI$^\dagger$ \end{tabular}
\\\hline15, QC & \begin{tabular}{c} $\epsilon_q=1$, $\epsilon_c=1$, $\eta_c=-1$, $\epsilon_{qc}=1$ \\ \hline $\epsilon_q=-1$, $\epsilon_c=1$, $\eta_c=-1$, $\epsilon_{qc}=1$ \end{tabular} & $\{\gamma ,Jm,C,JC\}_{Q}$ &  $\mathcal{R}_4$ & $\mathbb{Z}$ & $0$ & $\mathbb{Z}_2$ & $\mathbb{Z}_2$ & \begin{tabular}{c} AII \\ \hline NH CII$^\dagger$
\end{tabular}
\\\hline16, QC & \begin{tabular}{c} $\epsilon_q=1$,  $\epsilon_c=-1$, $\eta_c=1$, $\epsilon_{qc}=1$ \\ \hline $\epsilon_q=-1$,  $\epsilon_c=-1$, $\eta_c=1$, $\epsilon_{qc}=1$ \end{tabular} & $\{J\gamma ,m,C,JC\}_{Q}$ &  $\mathcal{R}_2$ & $\mathbb{Z}_2$ & $\mathbb{Z}_2$ & $\mathbb{Z}$ & $0$ & \begin{tabular}{c} D \\ \hline NH BDI \end{tabular}
\\\hline17, QC & \begin{tabular}{c} $\epsilon_q=1$, $\epsilon_c=-1$, $\eta_c=-1$, $\epsilon_{qc}=1$ \\ \hline $\epsilon_q=-1$,  $\epsilon_c=-1$, $\eta_c=-1$, $\epsilon_{qc}=1$ \end{tabular}  & $\{J\gamma ,m,C,JC\}_{Q}$ &  $\mathcal{R}_6$ & $0$ & $0$ & $\mathbb{Z}$ & $0$ & \begin{tabular}{c} C \\ \hline NH CII \end{tabular}
\\\hline18, QC & $\epsilon_c=1$, $\eta_c=1$, $\epsilon_{qc}=-1$ & $\{J\gamma ,m,\Sigma C,J\Sigma C\}_{Q}$ &  $\mathcal{R}_6$ & $0$ & $0$ & $\mathbb{Z}$ & $0$ & NH CI$^\dagger$
\\\hline19, QC & $\epsilon_c=1$, $\eta_c=-1$, $\epsilon_{qc}=-1$ & $\{J\gamma ,m,\Sigma C,J\Sigma C\}_{Q}$ &  $\mathcal{R}_2$ & $\mathbb{Z}_2$ & $\mathbb{Z}_2$ & $\mathbb{Z}$ & $0$ & NH DIII$^\dagger$
\\\hline20, QC & $\epsilon_c=-1$, $\eta_c=1$, $\epsilon_{qc}=-1$ & $\{\gamma ,Jm,\Sigma C,J\Sigma C\}_{Q}$ &  $\mathcal{R}_4$ & $\mathbb{Z}$ & $0$ & $\mathbb{Z}_2$ & $\mathbb{Z}_2$ & NH DIII
\\\hline21, QC & $\epsilon_c=-1$, $\eta_c=-1$, $\epsilon_{qc}=-1$ & $\{\gamma ,Jm,\Sigma C,J\Sigma C\}_{Q}$ &  $\mathcal{R}_0$ & $\mathbb{Z}$ & $0$ & $0$ & $0$ & NH CI \\\hline\end{tabular}\end{center}\caption{\label{ref:ptab1} First half of the periodic table with symmetries, commutation relations, Clifford algebra generators (a subscript indicates a commuting unitary symmetry, while a bracket indicates a unitary symmetry that squares to -1 and thus acts as an imaginary unit), classifying space, topological classification in low dimensions, as well as corresponding known classes. Note that depending on the commutation relations, equivalent representations can be physically distinct. To emphasize this, $\epsilon_q$ or $\epsilon_k$ is explicitly written when the value is important to distinguish different labels. }\end{table*}

\begin{table*}
\resizebox{2.05\columnwidth}{!}
{\begin{tabular}{ | c | c | c | c | c | c | c | c | c |}\hline Sym. & Gen. Rel. & Clifford Generators & Cl. Sp. 0D & $d = 0$ & 1 & 2 & 3 & Known Class\\\hline22, PQC & \begin{tabular}{c}$\epsilon_c=1$, $\eta_c=1$, $\epsilon_{pq}=1$, $\epsilon_{pc}=1$, $\epsilon_{qc}=1$\\\hline$\epsilon_c=-1$, $\eta_c=1$, $\epsilon_{pq}=1$, $\epsilon_{pc}=1$, $\epsilon_{qc}=1$\end{tabular} & \begin{tabular}{c}$\{\gamma ,Jm,JP,C,JC\}_{Q}$\\\hline$\{J\gamma ,m,JP,C,JC\}_{Q}$\end{tabular} &  $\mathcal{R}_1$ & $\mathbb{Z}_2$ & $\mathbb{Z}$ & $0$ & $0$ & BDI\\\hline23, PQC & \begin{tabular}{c}$\epsilon_c=1$, $\eta_c=-1$, $\epsilon_{pq}=1$, $\epsilon_{pc}=1$, $\epsilon_{qc}=1$\\\hline$\epsilon_c=-1$, $\eta_c=-1$, $\epsilon_{pq}=1$, $\epsilon_{pc}=1$, $\epsilon_{qc}=1$\end{tabular} & \begin{tabular}{c}$\{\gamma ,Jm,JP,C,JC\}_{Q}$\\\hline$\{J\gamma ,m,JP,C,JC\}_{Q}$\end{tabular} &  $\mathcal{R}_5$ & $0$ & $\mathbb{Z}$ & $0$ & $\mathbb{Z}_2$ & CII\\\hline24, PQC & \begin{tabular}{c}$\epsilon_c=1$, $\eta_c=1$, $\epsilon_{pq}=-1$, $\epsilon_{pc}=1$, $\epsilon_{qc}=1$\\\hline$\epsilon_c=-1$, $\eta_c=1$, $\epsilon_{pq}=-1$, $\epsilon_{pc}=1$, $\epsilon_{qc}=-1$\\\hline$\epsilon_c=1$, $\eta_c=1$, $\epsilon_{pq}=-1$, $\epsilon_{pc}=1$, $\epsilon_{qc}=-1$\\\hline$\epsilon_c=-1$, $\eta_c=1$, $\epsilon_{pq}=-1$, $\epsilon_{pc}=1$, $\epsilon_{qc}=1$\end{tabular} & \begin{tabular}{c}$\{\gamma ,Jm,C,JC\}_{Q}, [J\Sigma P]$\\\hline$\{\gamma ,Jm,\Sigma C,J\Sigma C\}_{Q}, [J\Sigma P]$\\\hline$\{J\gamma ,m,\Sigma C,J\Sigma C\}_{Q}, [J\Sigma P]$\\\hline$\{J\gamma ,m,C,JC\}_{Q}, [J\Sigma P]$\end{tabular} &  $\mathcal{C}_0$ & $\mathbb{Z}$ & $0$ & $\mathbb{Z}$ & $0$ & \\\hline25, PQC & \begin{tabular}{c}$\epsilon_c=1$, $\eta_c=-1$, $\epsilon_{pq}=-1$, $\epsilon_{pc}=1$, $\epsilon_{qc}=1$\\\hline$\epsilon_c=-1$, $\eta_c=-1$, $\epsilon_{pq}=-1$, $\epsilon_{pc}=1$, $\epsilon_{qc}=-1$\\\hline$\epsilon_c=1$, $\eta_c=-1$, $\epsilon_{pq}=-1$, $\epsilon_{pc}=1$, $\epsilon_{qc}=-1$\\\hline$\epsilon_c=-1$, $\eta_c=-1$, $\epsilon_{pq}=-1$, $\epsilon_{pc}=1$, $\epsilon_{qc}=1$\end{tabular} & \begin{tabular}{c}$\{\gamma ,Jm,C,JC\}_{Q}, [J\Sigma P]$\\\hline$\{\gamma ,Jm,\Sigma C,J\Sigma C\}_{Q}, [J\Sigma P]$\\\hline$\{J\gamma ,m,\Sigma C,J\Sigma C\}_{Q}, [J\Sigma P]$\\\hline$\{J\gamma ,m,C,JC\}_{Q}, [J\Sigma P]$\end{tabular} &  $\mathcal{C}_0$ & $\mathbb{Z}$ & $0$ & $\mathbb{Z}$ & $0$ & \\\hline26, PQC & \begin{tabular}{c}$\epsilon_c=1$, $\eta_c=1$, $\epsilon_{pq}=1$, $\epsilon_{pc}=-1$, $\epsilon_{qc}=1$\\\hline$\epsilon_c=-1$, $\eta_c=-1$, $\epsilon_{pq}=1$, $\epsilon_{pc}=-1$, $\epsilon_{qc}=1$\\\hline$\epsilon_c=1$, $\eta_c=1$, $\epsilon_{pq}=1$, $\epsilon_{pc}=-1$, $\epsilon_{qc}=-1$\\\hline$\epsilon_c=-1$, $\eta_c=-1$, $\epsilon_{pq}=1$, $\epsilon_{pc}=-1$, $\epsilon_{qc}=-1$\end{tabular} & \begin{tabular}{c}$\{\gamma ,Jm,P,C,JC\}_{Q}$\\\hline$\{J\gamma ,m,P,C,JC\}_{Q}$\\\hline$\{J\gamma ,m,P,\Sigma C,J\Sigma C\}_{Q}$\\\hline$\{\gamma ,Jm,P,\Sigma C,J\Sigma C\}_{Q}$\end{tabular} &  $\mathcal{R}_7$ & $0$ & $0$ & $0$ & $\mathbb{Z}$ & CI\\\hline27, PQC & \begin{tabular}{c}$\epsilon_c=1$, $\eta_c=-1$, $\epsilon_{pq}=1$, $\epsilon_{pc}=-1$, $\epsilon_{qc}=1$\\\hline$\epsilon_c=-1$, $\eta_c=1$, $\epsilon_{pq}=1$, $\epsilon_{pc}=-1$, $\epsilon_{qc}=1$\\\hline$\epsilon_c=1$, $\eta_c=-1$, $\epsilon_{pq}=1$, $\epsilon_{pc}=-1$, $\epsilon_{qc}=-1$\\\hline$\epsilon_c=-1$, $\eta_c=1$, $\epsilon_{pq}=1$, $\epsilon_{pc}=-1$, $\epsilon_{qc}=-1$\end{tabular} & \begin{tabular}{c}$\{\gamma ,Jm,P,C,JC\}_{Q}$\\\hline$\{J\gamma ,m,P,C,JC\}_{Q}$\\\hline$\{J\gamma ,m,P,\Sigma C,J\Sigma C\}_{Q}$\\\hline$\{\gamma ,Jm,P,\Sigma C,J\Sigma C\}_{Q}$\end{tabular} &  $\mathcal{R}_3$ & $0$ & $\mathbb{Z}_2$ & $\mathbb{Z}_2$ & $\mathbb{Z}$ & DIII\\\hline28, PQC & \begin{tabular}{c}$\epsilon_c=1$, $\eta_c=1$, $\epsilon_{pq}=-1$, $\epsilon_{pc}=-1$, $\epsilon_{qc}=1$\\\hline$\epsilon_c=-1$, $\eta_c=-1$, $\epsilon_{pq}=-1$, $\epsilon_{pc}=-1$, $\epsilon_{qc}=-1$\end{tabular} & \begin{tabular}{c}$\{\gamma ,Jm,C,JC\}_{Q,\Sigma P}$\\\hline$\{\gamma ,Jm,\Sigma C,J\Sigma C\}_{Q,\Sigma P}$\end{tabular} &  $\mathcal{R}_0^{\times 2}$ & $\mathbb{Z}^{\times 2}$ & $0$ & $0$ & $0$ & \\\hline29, PQC & \begin{tabular}{c}$\epsilon_c=1$, $\eta_c=-1$, $\epsilon_{pq}=-1$, $\epsilon_{pc}=-1$, $\epsilon_{qc}=1$\\\hline$\epsilon_c=-1$, $\eta_c=1$, $\epsilon_{pq}=-1$, $\epsilon_{pc}=-1$, $\epsilon_{qc}=-1$\end{tabular} & \begin{tabular}{c}$\{\gamma ,Jm,C,JC\}_{Q,\Sigma P}$\\\hline$\{\gamma ,Jm,\Sigma C,J\Sigma C\}_{Q,\Sigma P}$\end{tabular} &  $\mathcal{R}_4^{\times 2}$ & $\mathbb{Z}^{\times 2}$ & $0$ & $\mathbb{Z}_2^{\times 2}$ & $\mathbb{Z}_2^{\times 2}$ & \\\hline30, PQC & \begin{tabular}{c}$\epsilon_c=-1$, $\eta_c=1$, $\epsilon_{pq}=-1$, $\epsilon_{pc}=-1$, $\epsilon_{qc}=1$\\\hline$\epsilon_c=1$, $\eta_c=-1$, $\epsilon_{pq}=-1$, $\epsilon_{pc}=-1$, $\epsilon_{qc}=-1$\end{tabular} & \begin{tabular}{c}$\{J\gamma ,m,C,JC\}_{Q,\Sigma P}$\\\hline$\{J\gamma ,m,\Sigma C,J\Sigma C\}_{Q,\Sigma P}$\end{tabular} &  $\mathcal{R}_2^{\times 2}$ & $\mathbb{Z}_2^{\times 2}$ & $\mathbb{Z}_2^{\times 2}$ & $\mathbb{Z}^{\times 2}$ & $0$ & \\\hline31, PQC & \begin{tabular}{c}$\epsilon_c=-1$, $\eta_c=-1$, $\epsilon_{pq}=-1$, $\epsilon_{pc}=-1$, $\epsilon_{qc}=1$\\\hline$\epsilon_c=1$, $\eta_c=1$, $\epsilon_{pq}=-1$, $\epsilon_{pc}=-1$, $\epsilon_{qc}=-1$\end{tabular} & \begin{tabular}{c}$\{J\gamma ,m,C,JC\}_{Q,\Sigma P}$\\\hline$\{J\gamma ,m,\Sigma C,J\Sigma C\}_{Q,\Sigma P}$\end{tabular} &  $\mathcal{R}_6^{\times 2}$ & $0$ & $0$ & $\mathbb{Z}^{\times 2}$ & $0$ & \\\hline32, PQC & \begin{tabular}{c}$\epsilon_c=1$, $\eta_c=1$, $\epsilon_{pq}=1$, $\epsilon_{pc}=1$, $\epsilon_{qc}=-1$\\\hline$\epsilon_c=-1$, $\eta_c=1$, $\epsilon_{pq}=1$, $\epsilon_{pc}=1$, $\epsilon_{qc}=-1$\end{tabular} & \begin{tabular}{c}$\{J\gamma ,m,JP,\Sigma C,J\Sigma C\}_{Q}$\\\hline$\{\gamma ,Jm,JP,\Sigma C,J\Sigma C\}_{Q}$\end{tabular} &  $\mathcal{R}_5$ & $0$ & $\mathbb{Z}$ & $0$ & $\mathbb{Z}_2$ & \\\hline33, PQC & \begin{tabular}{c}$\epsilon_c=1$, $\eta_c=-1$, $\epsilon_{pq}=1$, $\epsilon_{pc}=1$, $\epsilon_{qc}=-1$\\\hline$\epsilon_c=-1$, $\eta_c=-1$, $\epsilon_{pq}=1$, $\epsilon_{pc}=1$, $\epsilon_{qc}=-1$\end{tabular} & \begin{tabular}{c}$\{J\gamma ,m,JP,\Sigma C,J\Sigma C\}_{Q}$\\\hline$\{\gamma ,Jm,JP,\Sigma C,J\Sigma C\}_{Q}$\end{tabular} &  $\mathcal{R}_1$ & $\mathbb{Z}_2$ & $\mathbb{Z}$ & $0$ & $0$ & \\\hline34, K & \begin{tabular}{c} $\epsilon_k=1$, $\eta_k=1$ \\ \hline $\epsilon_k=-1$,  $\eta_k=1$ \end{tabular}  & $\{\gamma ,Jm,J\Sigma ,K,JK\}$ &  $\mathcal{R}_1$ & $\mathbb{Z}_2$ & $\mathbb{Z}$ & $0$ & $0$ & \begin{tabular}{c} NH AI \\ \hline NH D$^\dagger$ \end{tabular} \\\hline35, K & \begin{tabular}{c} $\epsilon_k=1$, $\eta_k=-1$ \\ \hline $\epsilon_k=-1$,  $\eta_k=-1$ \end{tabular} & $\{\gamma ,Jm,J\Sigma ,K,JK\}$ &  $\mathcal{R}_5$ & $0$ & $\mathbb{Z}$ & $0$ & $\mathbb{Z}_2$ & \begin{tabular}{c} NH AII \\ \hline NH C$^\dagger$ \end{tabular} \\\hline36, PK & $\eta_k=1$, $\epsilon_{pk}=1$ & $\{\gamma ,Jm,J\Sigma ,K,JK\}_{\Sigma P}$ &  $\mathcal{R}_1^{\times 2}$ & $\mathbb{Z}_2^{\times 2}$ & $\mathbb{Z}^{\times 2}$ & $0$ & $0$ & \\\hline37, PK & $\eta_k=-1$, $\epsilon_{pk}=1$ & $\{\gamma ,Jm,J\Sigma ,K,JK\}_{\Sigma P}$ &  $\mathcal{R}_5^{\times 2}$ & $0$ & $\mathbb{Z}^{\times 2}$ & $0$ & $\mathbb{Z}_2^{\times 2}$ & \\\hline38, PK & \begin{tabular}{c}$\eta_k=1$, $\epsilon_{pk}=-1$\\\hline$\eta_k=-1$, $\epsilon_{pk}=-1$\end{tabular} & \begin{tabular}{c}$\{\gamma ,Jm,J\Sigma ,K,JK\}, [J\Sigma P]$\\\hline$\{\gamma ,Jm,J\Sigma ,K,JK\}, [J\Sigma P]$\end{tabular} &  $\mathcal{C}_1$ & $0$ & $\mathbb{Z}$ & $0$ & $\mathbb{Z}$ &\\\hline\end{tabular}
}
\caption{\label{ref:ptab2} Second half of the periodic table (continued from the previous one). }
\end{table*}

\end{document}